\documentclass[12pt]{article}

\usepackage[dvips]{graphicx}

\makeatletter
\def\@cite#1#2{{[{#1}]\if@tempswa\typeout
{IJCGA warning: optional citation argument
ignored: `#2'} \fi}}

\newcount\@tempcntc
\def\@citex[#1]#2{\if@filesw\immediate\write\@auxout{\string\citation{#2}}\fi
  \@tempcnta\z@\@tempcntb\m@ne\def\@citea{}\@cite{\@for\@citeb:=#2\do
    {\@ifundefined
       {b@\@citeb}{\@citeo\@tempcntb\m@ne\@citea\def\@citea{,}{\bf ?}\@warning
       {Citation `\@citeb' on page \thepage \space undefined}}%
    {\setbox\z@\hbox{\global\@tempcntc0\csname b@\@citeb\endcsname\relax}%
     \ifnum\@tempcntc=\z@ \@citeo\@tempcntb\m@ne
       \@citea\def\@citea{,}\hbox{\csname b@\@citeb\endcsname}%
     \else
      \advance\@tempcntb\@ne
      \ifnum\@tempcntb=\@tempcntc
      \else\advance\@tempcntb\m@ne\@citeo
      \@tempcnta\@tempcntc\@tempcntb\@tempcntc\fi\fi}}\@citeo}{#1}}
\def\@citeo{\ifnum\@tempcnta>\@tempcntb\else\@citea\def\@citea{,}%
  \ifnum\@tempcnta=\@tempcntb\the\@tempcnta\else
   {\advance\@tempcnta\@ne\ifnum\@tempcnta=\@tempcntb \else \def\@citea{--}\fi
    \advance\@tempcnta\m@ne\the\@tempcnta\@citea\the\@tempcntb}\fi\fi}
\makeatother
\newenvironment{Eqnarray}%
     {\arraycolsep 0.14em\begin{eqnarray}}{\end{eqnarray}}

\def\be{\begin{equation}}
\def\ee{\end{equation}}
\def\bear{\be\begin{array}}
\def\eear{\end{array}\ee}
\def\bea{\begin{Eqnarray}}
\def\eea{\end{Eqnarray}}

\def\lsim{\mathrel{\raise.3ex\hbox{$<$\kern-.75em\lower1ex\hbox{$\sim$}}}}
\def\gsim{\mathrel{\raise.3ex\hbox{$>$\kern-.75em\lower1ex\hbox{$\sim$}}}}
\def\ifmath#1{\relax\ifmmode #1\else $#1$\fi}
\def\ls#1{\ifmath{_{\lower1.5pt\hbox{$\scriptstyle #1$}}}}

\def\beq{\begin{equation}}
\def\eeq{\end{equation}}
\def\beqa{\begin{Eqnarray}}
\def\eeqa{\end{Eqnarray}}

\def\boxit#1{\leavevmode\thinspace\hbox{\vrule\vtop{\vbox{\hrule%
        \vskip3pt\kern1pt\hbox{\vphantom{\bf/}\thinspace\thinspace%
        {\bf#1}\thinspace\thinspace}}\kern1pt\vskip3pt\hrule}\vrule}%
        \thinspace}
\def\Boxit#1{\noindent\vbox{\hrule\hbox{\vrule\kern3pt\vbox{
        \advance\hsize-7pt\vskip-\parskip\kern3pt\bf#1
        \hbox{\vrule height0pt depth\dp\strutbox width0pt}
        \kern3pt}\kern3pt\vrule}\hrule}}

\def\baselinestretch{1}
\begin{document}

\catcode`@=11
\newtoks\@stequation
\def\subequations{\refstepcounter{equation}%
\edef\@savedequation{\the\c@equation}%
  \@stequation=\expandafter{\theequation}
  \edef\@savedtheequation{\the\@stequation}
  \edef\oldtheequation{\theequation}%
  \setcounter{equation}{0}%
  \def\theequation{\oldtheequation\alph{equation}}}
\def\endsubequations{\setcounter{equation}{\@savedequation}%
  \@stequation=\expandafter{\@savedtheequation}%
  \edef\theequation{\the\@stequation}\global\@ignoretrue
\noindent}
\catcode`@=12
\setcounter{footnote}{1} \setcounter{page}{1}

\noindent

\title{{\bf Reconstructing the two right-handed neutrino model}}
\vskip2in
\author{    
{\bf Alejandro Ibarra\footnote{\baselineskip=16pt  E-mail: {\tt
alejandro.ibarra@cern.ch}}} \\ 
\hspace{3cm}\\
{\small Instituto de F\'{\i}sica Te\'orica, CSIC/UAM, C-XVI} \\
{\small Universidad Aut\'onoma de Madrid,} \\
{\small Cantoblanco, 28049 Madrid, Spain.}
}
\maketitle

\def\baselinestretch{1.15}
\begin{abstract}
\noindent
In this paper we propose a low-energy parametrization
of the two right-handed neutrino model, and discuss
the prospects to determine experimentally these parameters 
in supersymmetric scenarios. In addition,
we present exact formulas to reconstruct
the high-energy leptonic superpotential in 
terms of the low-energy observables.
We also discuss limits of the three right-handed
neutrino model where this procedure applies.
\end{abstract}

\thispagestyle{empty}

\vskip-14.5cm
\rightline{IFT-UAM/CSIC-05-46}

\newpage
\baselineskip=20pt

\section{Introduction}

The observed neutrino masses can be naturally accounted for
by adding to the Standard Model Lagrangian three heavy
singlets, usually identified with the right-handed neutrinos.
This framework is denoted as the type I
see-saw mechanism, or simply the see-saw mechanism \cite{seesaw}.
Furthermore, this framework can also accommodate, although
not explain, the large atmospheric and solar mixing angles, 
as well as the small 13 element in the leptonic mixing matrix. 

Although the see-saw mechanism describes qualitatively well
the observations, it lacks predictive power. The reason
is that the leptonic Lagrangian is defined at high energies
by 21 parameters, whereas experimentally we can only measure
at most 12 parameters. The 9 remaining parameters are lost
in the decoupling process and can be arbitrarily chosen, 
without leaving any other trace at low energies.\footnote{
There is additional information encoded in dimension six operators,
but unfortunately they are too suppressed to be observed experimentally
\cite{Broncano:2002rw}.}

In the supersymmetric version of the see-saw mechanism
the neutrino Yukawa coupling affects the renormalization
group evolution of the slepton parameters above the
decoupling scale \cite{Borzumati:1986qx}, 
and thus leave an imprint that could
be disentangled using low energy experiments (after making some assumptions
on the structure of the slepton parameters at the cut-off
scale). For instance, phenomena such as rare decays
\cite{Hisano:1995nq,Casas:2001sr,Lavignac:2001vp}
electric dipole moments \cite{Romanino:2001zf}, 
or mass splittings among the  different generations 
of charged sleptons or sneutrinos \cite{Baer:2000hx} 
provide additional information about the see-saw parameters. 

This information is encoded in a $3 \times 3$ hermitian 
matrix, $P={\bf Y}^{\dagger}_{\nu}{\bf Y}_{\nu}$, that
depends on nine parameters (six moduli and three phases).
It can be shown that these nine parameters
are precisely the complementary information needed
to reconstruct the high energy Lagrangian
 \cite{Davidson:2001zk,Ellis:2002fe}. The positive 
consequence of this observation is that the see-saw mechanism can
be parametrized just in terms of low energy observables;
the negative consequence is that for any set of low
energy lepton and slepton parameters, there is a
high energy theory with three right-handed neutrinos that
can accommodate it, and in consequence, the see-saw model
cannot be disproved.

In this parametrization, the correspondence between high energy
and low energy parameters is one to one. Therefore, any
additional hypothesis on the high-energy see-saw parameters
would lead to predictions on the low energy parameters.
Some well motivated assumptions have been proposed 
in the literature, such as texture zeros, symmetric matrices
or the two right-handed neutrino model, and their consequences
studied in a number of papers.
In this work we will concentrate on the latter 
possibility \cite{2RHN,related}. The motivation is the following.
Leaving aside the LSND anomaly, experiments have measured
two mass splittings, indicating that at least two new
scales have to be introduced. The two new scales could
correspond to the masses of two right handed neutrinos, being
the third one not necessary to reproduce the 
oscillation experiments. Therefore, a model with just two
right handed neutrinos could explain all the observations,
but it depends on less parameters.

The two right-handed neutrino
model has some interesting features. Namely, the two right-handed
neutrino model predicts a hierarchical spectrum for the light
neutrinos, being the lightest strictly massless.
Furthermore, since there are only
two non-vanishing masses, there is only one Majorana phase,
corresponding to the phase difference between these two eigenvalues.
Analogously to the model with three right-handed neutrinos,
there are three mixing angles and one Dirac phase.

Since the number of parameters involved is smaller, 
we expect predictions not only for the neutrino mass matrix,
but also for the low-energy slepton mass matrix
(under some assumptions about the structure of this matrix
at high energy),
as well as a simpler reconstruction procedure of the high 
energy parameters. In section 2 we will show
that in the two right-handed neutrino model some
relations arise among different elements in the
slepton mass matrix. We will also present a possible parametrization
of this model just in terms of low energy observables. In
Section 3 we will discuss the prospects to determine experimentally
these parameters, and accordingly the feasibility of 
the reconstruction of the high-energy parameters. In section
4 we will propose an exact reconstruction procedure, and illustrate it
for a particular possibility of the low energy parameters.
In section 5 we will analyze different limits of the three right
handed neutrino model that can be well described in practice
by a two right-handed neutrino model, and where the reconstruction
procedure proposed in this paper applies. Finally, in 
Section 6 we will present our conclusions. We will also present an appendix 
with a more elaborated reconstruction procedure.

\section{Parametrizations of the two right-handed neutrino model}

In the appropriate basis, the two right-handed neutrino 
(2RHN) model is defined at high energies by a $2\times 3$ Yukawa
matrix and two right-handed neutrino masses, $M_1$ and $M_2$.
This amounts to eight moduli and three phases. On the other hand, at low
energies the neutrino mass matrix is defined by five moduli
(two masses and three mixing angles) plus two phases (the Dirac phase
and the Majorana phase). Therefore, the number of unknown parameters
is reduced to three moduli and one phase.

Using the parametrization presented in \cite{Casas:2001sr},
the neutrino Yukawa coupling can be expressed as:
\bea
{\bf Y}_{\nu}=D_{\sqrt{{\bf M_{\nu}}}}R D_{\sqrt{m}}U^{\dagger }/\langle
H_u^0\rangle, 
\label{yukawa}
\eea
where $D_{\sqrt{{\bf M_{\nu}}}}={\rm diag}(\sqrt{M_1},\sqrt{M_2})$ 
is the diagonal
matrix of the square roots of the right-handed masses,
$D_{\sqrt{m}}={\rm diag}(0,\sqrt{m_2},\sqrt{m_3})$ 
is the diagonal matrix of the squared
roots of the physical masses of the light neutrinos, 
$\langle H_u^0\rangle$ is the vacuum expectation value
of the neutral component of the up-type Higgs doublet, $U$ is the
leptonic mixing matrix \cite{Maki:mu}, and $R$ is a $2\times 3$ 
complex matrix, which parametrizes the information
that is lost in the decoupling of the right-handed neutrinos. 
It is possible to prove that $R$ has the following structure 
\cite{Ibarra:2003xp,Ibarra:2003up}
\bea
R=\pmatrix{
0 & \cos z & \xi \sin z \cr
0 & -\sin z & \xi \cos z},
\label{R2x3}
\eea
where $z$ is a complex parameter and $\xi=\pm 1$ 
is a parameter that accounts for a 
discrete indeterminacy in the Yukawa coupling.

Notice that we have included all the low
energy phases in the definition of the matrix $U$, 
{\it i.e.} we have written the leptonic mixing matrix in the 
form $U=V~{\rm diag}~(1, e^{-i\phi/2},1) $,
where $\phi $ is  the Majorana phase and $V$ has
the form of the CKM matrix:
\bea
V=\pmatrix{c_{13}c_{12} & c_{13}s_{12} & s_{13}e^{-i\delta}\cr
-c_{23}s_{12}-s_{23}s_{13}c_{12}e^{i\delta} & c_{23}c_{12}-s_{23}s_{13}s_{12}e^{i\delta} & s_{23}c_{13}\cr
s_{23}s_{12}-c_{23}s_{13}c_{12}e^{i\delta} & -s_{23}c_{12}-c_{23}s_{13}s_{12}e^{i\delta} &
c_{23}c_{13}\cr},
\label{Vdef}  
\eea
so that the neutrino mass matrix is 
${\cal M}=U^* {\rm diag}(0,m_2,m_3) U^{\dagger}$.
It is straightforward to check that eq.(\ref{yukawa})
indeed satisfies ${\cal M}= {\bf Y}_{\nu}^T {\rm diag}(M^{-1}_1,M^{-1}_2) 
{\bf Y}_{\nu}\langle H_u^0\rangle^2$.

The Yukawa coupling affects the renormalization group
equation of the slepton parameters 
through the combination $P={\bf Y}^{\dagger}_{\nu}{\bf Y}_{\nu}$,
that depends in general on six moduli and three phases.
Since the Yukawa coupling depends in the 2RHN model 
on only three unknown
moduli and one phase, so does $P$, and consequently
it is possible to obtain predictions on the moduli
of three $P$-matrix elements and the phases of two 
$P$-matrix elements. Namely, from eq.(\ref{yukawa})
one obtains that:
\bea
U^{\dagger} P U =  U^{\dagger} {\bf Y}^{\dagger}_{\nu}{\bf Y}_{\nu} U = 
D_{\sqrt{m}} R^{\dagger} D_{{\bf M_{\nu}}} R D_{\sqrt{m}}
/\langle H^0\rangle^2.
\label{mastereq}
\eea
Since $m_1=0$, it follows that $(U^{\dagger} P U)_{1i}=0$, for $i=1,2,3$,
leading to three relations among the elements in $P$. For instance,
one could derive the diagonal elements in $P$ in terms of the 
off-diagonal elements:
\bea
P_{11}&=&-\frac{P^*_{12}U^*_{21}+P^*_{13}U^*_{31}}
{U^*_{11}}, \nonumber \\
P_{22}&=&-\frac{P_{12}U^*_{11}+P^*_{23}U^*_{31}}{U^*_{21}}, \nonumber \\ 
P_{33}&=&-\frac{P_{13}U^*_{11}+P_{23}U^*_{21}}{U^*_{31}}.
\label{diagonal}
\eea
The observation of these correlations would be non-trivial
tests of the 2RHN model.

The relations for the phases arise from the hermicity
of $P$, since the diagonal elements in $P$ have
to be real. Taking as the independent phase the argument
of $P_{12}$, one can derive from eq.(\ref{diagonal})
the arguments of the remaining elements:
\bea
e^{i{\rm arg}P_{13}}&=&\frac{-i~{\rm Im}(P_{12}U_{21}U^*_{11})\pm
\sqrt{|P_{13}|^2|U_{11}|^2|U_{31}|^2-[{\rm Im}(P_{12}U_{21}U^*_{11})]^2}}
{ |P_{13}| U_{31}U^*_{11}},
\nonumber \\
e^{i{\rm arg}P_{23}}&=&\frac{i~{\rm Im}(P_{12}U_{21}U^*_{11})\pm
\sqrt{|P_{23}|^2|U_{21}|^2|U_{31}|^2-[{\rm Im}(P_{12}U_{21}U^*_{11})]^2}}
{ |P_{23}| U_{31}U^*_{21}},
\label{phases}
\eea
where the $\pm$ sign has to be chosen so that
the eigenvalues of $P$ are positive. It is important to
remark that the hermicity of $P$ is not guaranteed for any value
of $P_{12}$, $|P_{13}|$, $|P_{23}|$; only some
particular ranges for the parameters are allowed, corresponding to the
values for which the arguments of the square roots in 
eq.(\ref{phases}) are positive.

We conclude then that the $P$-matrix parameters 
$P_{12}$, $|P_{13}|$ and $|P_{23}|$
can be regarded as independent and can 
be used as an alternative parametrization of the 2RHN model.
Together with the five moduli and the two phases of the neutrino
mass matrix, sum up to the eight moduli and the three phases
necessary to reconstruct the high-energy Lagrangian of the
2RHN model.

\section{Can we reconstruct (realistically) the complete theory
from low energy observables?}

An interesting feature of the 2RHN model is that
it could be feasible to reconstruct the
neutrino mass matrix. This model predicts that one of the 
neutrino masses vanishes, and in consequence the spectrum is necessarily 
hierarchical. In this scenario, the masses
would have been already determined by present oscillation experiments:
$m_2=\sqrt{\Delta m^2_{21}}$ and $m_3=\sqrt{\Delta m^2_{31}}$,
where $\Delta m^2_{21}=(7.1-8.9)\times 10^{-5}$eV and
$\Delta m^2_{31}=(1.4-3.3)\times 10^{-3}$eV are
the $3\sigma$ ranges for the solar and atmospheric 
mass splittings obtained from the combined 
analysis of global data  \cite{Maltoni:2004ei}.
In the next few years, the measurements of the mass splittings
are expected to improve. To be specific, the error in the measurement
of the atmospheric mass splitting, $\Delta m^2_{31}$,
is expected to be reduced by experiments using 
the CERN to Gran Sasso neutrino beam 
(ICARUS \cite{Aprili:2002wx} and OPERA \cite{Guler:2000bd}), 
MINOS \cite{Ables:1995wq},
NO$\nu$A \cite{Ayres:2004js} and particularly T2K \cite{Itow:2001ee},
that will probably reduce the present error by one 
order of magnitude. There are also proposals to reduce the error
in the measurement of the solar mass splitting, $\Delta m^2_{21}$.
Namely, if the SuperK detector was loaded with gadolinium, it would
be possible to reduce the error by a factor of six 
\cite{Choubey:2004bf}.

Concerning the mixing angles, two of them have been determined
to a good accuracy by present experiments: $\sin^2 {\theta_{12}}=
0.24-0.40$ and $\sin^2 {\theta_{23}}=0.34-0.68$ at the
$3\sigma$ level \cite{Maltoni:2004ei}. The T2K experiment will probably
reduce the error in  $\sin^2 {\theta_{23}}$ by a factor of two.
On the other hand, the error in $\sin^2 {\theta_{12}}$ will not
be substantially reduced in the near future, although an experiment
similar to KamLAND but with a baseline slightly shorter,
$L\sim60$km, could reduce the error by a factor of four 
\cite{Bandyopadhyay:2004cp}.

The angle $\theta_{13}$ has not
been detected yet but the global analysis sets the upper bound
$\sin^2\theta_{13}<0.046$, also
at $3\sigma$ \cite{Maltoni:2004ei}. Ongoing experiments such as MINOS, ICARUS
or OPERA could improve the present limit by a factor of two,
while future experiments as D-Chooz \cite{Ardellier:2004ui}
 by a factor of four, and T2K or NO$\nu$A by a factor of ten. 
If $\theta_{13}$ is large, the combined analysis 
of the experiments could provide
some information about the Dirac phase $\delta$. However,
the detailed analysis of CP violation in the neutrino sector
will require superbeams. For instance, improving the proton
intensity at JHF to 4MW and using the proposed Hyper-Kamiokande
detector, it could be possible to reach a sensitivity below
$3\times 10^{-4}$ for $\sin^2\theta_{13}$ and around $10^{\circ}-20^{\circ}$
for $\delta$ \cite{Itow:2001ee}.

Thus we find that there are good prospects to determine the
masses, mixing angles and $\delta$ to the percent
level in the next 10-20 years \cite{hernandez}. However,
the measurement of the Majorana phase is still very challenging.
Although it could in principle be measured in 
experiments of neutrinoless double beta decay,
the sensitivity of the projected experiments
is still far from observing this process in the case of
a hierarchical spectrum of neutrinos \cite{Pascoli:2005zb}.

Concerning the matrix $P$, there are also very good prospects
to determine the relevant parameters: $P_{12}$, $|P_{13}|$ and $|P_{23}|$.
With experiments of rare decays, it would be possible
to determine the moduli of the off-diagonal elements of $P$.
The present bounds on the rates for these decays are 
$BR(\mu\rightarrow e\gamma) \lsim1.2\times10^{-11}$ \cite{Brooks:1999pu},
$BR(\tau\rightarrow \mu\gamma) < 3.1\times 10^{-7}$ \cite{Abe:2003sx}
(Belle) or
$<6.8\times10^{-8}$ \cite{Aubert:2005ye} (BaBar), and 
$BR(\tau\rightarrow e\gamma) <3.9\times 10^{-7}$ \cite{Hayasaka:2005xw},
that translate into $|P_{12}|<3\times10^{-4}$, $|P_{23}|<0.09$ 
and $|P_{13}|<0.09$, for $\tan\beta=10$
and typical slepton masses of $\sim 400$GeV \cite{Masiero:2004js}.

In the next few years, the sensitivity of the experiments 
to these processes is also expected to improve. 
The  MEG experiment expects to reach a sensitivity
of $BR(\mu\rightarrow e\gamma)\lsim 10^{-13}$
\cite{Baldini:2004dj},
that would translate into $|P_{12}|\lsim3\times10^{-5}$. 
Although presently the most stringent
constraints on $|P_{12}|$ arise from the process 
$\mu\rightarrow e\gamma$, this role could be
played in the future by experiments on the 
coherent conversion of muons into electrons in nuclei.
\footnote{When the photon penguin diagram dominates the 
contribution in the $\mu-e$ conversion, the conversion
rate is approximately three orders of magnitude 
smaller than the branching ratio for $\mu\rightarrow
e \gamma$.}
The current experimental bounds are $R(\mu^-{\rm Ti}\rightarrow
e^- {\rm Ti})<6.1\times10^{-13}$ \cite{Wintz} and
$R(\mu^-{\rm Au}\rightarrow
e^- {\rm Au})<5\times10^{-13}$ \cite{vanderSchaaf:2003ar},
and are expected to be improved 
by the PRISM/PRIME experiment, aiming to a sensitivity
of $10^{-18}$ \cite{Sato}, or the  CERN neutrino factory,
aiming to $10^{-19}$ \cite{Aysto:2001zs}.

On the other hand, B-factories are also $\tau$-factories,
and constitute splendid opportunities to search for
rare $\tau$ decays. Future super B-factories could
produce of the order of $10^{10}$ $\tau$ pairs
at a luminosity of 10 $ab^{-1}$, allowing to probe
branching ratios  for the rare $\tau$ decays down
to the level of $10^{-8}-10^{-9}$ \cite{Akeroyd:2004mj}.

The only phase that appears in $P$ could be determined
from measurements of the electric dipole moment of the electron,
whose present bound is $d_e<1.6\times 10^{-27} {\rm ~e~cm}$ 
\cite{Regan:2002ta}. In order
to disentangle the contribution from the neutrino Yukawa
couplings it would be necessary to determine the phases
in the neutralino or the chargino sector from other experiments, most
probably the LHC or the ILC. The prospects to improve the
sensitivity of the experiments to detect electron dipole moments
are also very encouraging.
By using the metastable $a(1)[^3\Sigma^*]$ state of PbO
it could be possible to improve the sensitivity
of the experiments to $d_e\lsim 10^{-29}{\rm ~e~cm}$
\cite{Kawall:2003ga}, or even to  
$d_e\lsim 10^{-31}{\rm ~e~cm}$ in a few years
\cite{Kawall:2004nv}. In the longer term, it could
be possible to improve the sensitivity to 
$d_e\lsim 10^{-35}{\rm ~e~cm}$ using solid state 
techniques \cite{Lamoreaux:2001hb}. Incidentally,
it has been argued that  the effect of the phases
in the neutrino Yukawa coupling would not be observed
in the electron dipole moment before this sensitivity is
reached \cite{Demir:2005ya}.

In summary, the prospects to detect or further constrain
the low-energy parameters of the 2RHN model are very encouraging,
and accordingly the prospects to reconstruct the high
energy theory from experiments. It is remarkable that 
most of the problems encountered when determining
the low energy parameters of the 3RHN model disappear
in the 2RHN model. Namely, in the 3RHN model there are two Majorana
phases, and despite one combination of them could be measured
in experiments of neutrinoless double beta decay,
there is no proposed experimental set-up to measure the second combination.
Furthermore, in the matrix $P$ there are more 
independent parameters in the 3RHN model than in the 2RHN model,
 and the prospects to measure them are not 
so encouraging. In particular, the diagonal
elements in $P$ would be quite hard to measure, although it could
be possible to measure  the largest mass difference among 
the sleptons or the sneutrinos in colliders \cite{Baer:2000hx}.
On the other hand, the measurement of the 
smallest mass difference seems to be far out of the 
reach of the proposed future experiments.
Concerning the electric dipole moments, there are some prospects
to improve the present bound on the muon electric dipole moment,
$d_{\mu}<7\times 10^{-19}{\rm~e~cm}$ \cite{Harris:1999jx}, to 
$d_{\mu}\lsim 10^{-24}{\rm~e~cm}$ using the muon ring
at BNL \cite{Semertzidis:1999kv} or even to 
$d_{\mu}\lsim 10^{-26}{\rm~e~cm}$ at the neutrino 
factory \cite{Aysto:2001zs}. On the other hand, the present bounds on
the $\tau$ electric dipole moment are very loose,
$-2.2<{\rm Re}(d_{\tau})<4.5(\times10^{-17}){\rm~e~cm}$ \cite{Inami:2002ah},
and there are no prospects to improve them substantially
in the near future.

It was pointed out in \cite{Davidson:2002qv,Chankowski:2003rr,Ibarra:2003up}
that the phase of $z$ is the only
phase that plays a role in the mechanism of
leptogenesis \cite{Fukugita:1986hr}. Therefore,
the see-saw mechanism could be parametrized in terms of the
leptogenesis phase instead of the phase that induces
electric dipole moments. Nevertheless, the leptogenesis
mechanism depends on assumptions that are harder
to test than the assumptions needed to disentangle the phase
in $P$ from electric dipole moments, and thus this
possibility does not seem to be very practical. In any case,
we find very remarkable the close relation between
electric dipole moments and leptogenesis in the
2RHN model.

\section{The reconstruction procedure}

In Section 2 we have discussed that the 
complete Lagrangian can be written in terms of the 
five moduli and two phases of the neutrino mass matrix,
and the three independent moduli and two phases
of the matrix $P$, that is involved in the
radiative corrections of the slepton parameters. In this
section we will derive {\it exact} formulas for
the high energy parameters in terms of these low energy
parameters \cite{previous}

To this end, we will use the parametrization of the Yukawa
couplings in eq.(\ref{yukawa}), so that all our ignorance
of the high energy theory is encoded in the right-handed
neutrino masses, $M_1$ and $M_2$, and the complex parameter
in the matrix $R$, $z$.
Let us define the hermitian matrix  $Q\equiv U^{\dagger} P U$, 
that depends {\it exclusively} on parameters that in principle
could be measured in low energy experiments. The first 
row and column {vanish and yield the relations among the $P$-matrix
elements already presented in eq.(\ref{diagonal}). On the other hand,
the remaining elements $Q_{22}$, $Q_{23}$,  $Q_{33}$, 
can be written in terms of the high-energy parameters $M_1$,
$M_2$ and $z$. Therefore, one can invert the equations to
derive exact expressions for the high-energy parameters
in terms of the low energy parameters in $Q$. These
expressions are:
\bea
M_1&=&\frac{1}{2}\left[
\sqrt{\left(\frac{Q_{33}}{ m_3}+\frac{Q_{22}}{m_2}\right)^2+
\frac{(Q_{23}-Q^*_{23})^2}{ m_2 m_3}}
-\sqrt{\left(\frac{Q_{33}}{ m_3}-\frac{Q_{22}}{m_2}\right)^2+
\frac{(Q_{23}+Q^*_{23})^2}{ m_2 m_3}}\right] \langle H_u^0 \rangle^2, 
\nonumber \\
M_2&=&\frac{1}{2}\left[
\sqrt{\left(\frac{Q_{33}}{ m_3}+\frac{Q_{22}}{m_2}\right)^2+
\frac{(Q_{23}-Q^*_{23})^2}{ m_2 m_3}}
+\sqrt{\left(\frac{Q_{33}}{ m_3}-\frac{Q_{22}}{m_2}\right)^2+
\frac{(Q_{23}+Q^*_{23})^2}{ m_2 m_3}}\right]\langle H_u^0 \rangle^2 ,
\nonumber \\
\cos2z&=&
\left(\frac{Q^2_{33}}{m^2_3}-\frac{Q^2_{22}}{m^2_2}+
\frac{(Q_{23}+Q^*_{23})(Q_{23}-Q^*_{23})}{m_2 m_3}\right)
\frac{\langle H_u^0 \rangle^4}{M^2_2-M^2_1}.
\label{reconstruction}
\eea
To complete the reconstruction procedure, the Yukawa
coupling would be derived from these parameters
using eq.(\ref{yukawa}) and where the discrete
parameter $\xi$ in eq.(\ref{R2x3}) is determined by:
\bea
\xi=\frac{\sqrt{m_2 m_3}}{Q_{23}\langle H_u^0 \rangle^2}
(M_1 \sin z \cos z^*-M_2 \cos z \sin z^*).
\eea

In the case that all the parameters are real, 
the expressions greatly simplify: 
\bea
M_1&=&\frac{1}{2}\left[\frac{Q_{33}}{m_3}+\frac{Q_{22}}{ m_2}
-\sqrt{\left(\frac{Q_{33}}{ m_3}-\frac{Q_{22}}{m_2}\right)^2+
\frac{4Q^2_{23}}{ m_2 m_3}}\right]\langle H_u^0 \rangle^2,
\nonumber \\
M_2&=&\frac{1}{2}\left[\frac{Q_{33}}{m_3}+\frac{Q_{22}}{ m_2}
+\sqrt{\left(\frac{Q_{33}}{ m_3}-\frac{Q_{22}}{m_2}\right)^2+
\frac{4Q^2_{23}}{ m_2 m_3}}\right]\langle H_u^0 \rangle^2,
\nonumber \\
\cos2z&=&\left(\frac{Q^2_{33}}{m^2_3}-\frac{Q^2_{22}}{m^2_2}\right)
\frac{\langle H_u^0 \rangle^4}{M^2_2-M^2_1}.
\label{reconstruction-real}
\eea

We would like to illustrate now the reconstruction
procedure for an interesting possibility for the matrix $P$.
In the previous section we argued that among all the elements
in $P$, the ones with better prospects to be constrained
or measured were the off-diagonal ones. Therefore, it will
prove convenient from the phenomenological point of view
to use $P_{12}$, $|P_{13}|$ and $|P_{23}|$, 
together with the neutrino mass matrix, to parametrize
the 2RHN model. Furthermore, the stringent constraints 
on the process $\mu \rightarrow e \gamma$ suggests us 
to take the limit $|P_{12}|\ll  |P_{13}|, 
|P_{23}|$.\footnote{It is important to stress 
that there is no solid experimental evidence supporting
this possibility. The only reason why the bound
on $\mu\rightarrow e \gamma$ is stronger than the bounds on
$\tau\rightarrow (\mu, e) \gamma$ is that
presently the  muon sources are more intense than the
tau sources. Other possibilities are a priori
equally plausible from the phenomenological point
of view, despite there is a theoretical
prejudice in favor of $|P_{12}|\ll  |P_{13}|, 
|P_{23}|$. The analysis for other scenarios will be presented
elsewhere.} This limit yields a very constrained
structure for the $P$-matrix,
\bea
P\simeq
\pmatrix{
-P^*_{13}\frac{U^*_{31}}{U^*_{11}}& 0 & P_{13} \cr
0 & -P^*_{23} \frac{U^*_{31}}{U^*_{21}} & P_{23} \cr
P^*_{13} & P^*_{23} & -P_{13}\frac{U^*_{11}}{U^*_{31}}
-P_{23}\frac{U^*_{21}}{U^*_{31}}},
\label{P-interesting}
\eea
that in turn would imply predictions for the 
mass splittings among the sleptons in terms
of the branching ratios for the processes
$\tau\rightarrow \mu \gamma$ and
$\tau\rightarrow e \gamma$. Note also
that the phases in the matrix $P$ are
fixed in terms of the phases in the leptonic
mixing matrix. Namely, to make the diagonal
elements of the matrix $P$ real, as required by
hermicity, one has to require
\bea
{\rm arg}P_{13}&\simeq& {\rm arg} U_{11} -{\rm arg} U_{31} +\pi,
\nonumber \\
{\rm arg}P_{23}&\simeq& {\rm arg} U_{21} -{\rm arg} U_{31} +\pi.
\label{arguments-interesting}
\eea
We have also resolved the $\pm$ ambiguity in the expressions
for the arguments of  $P_{13}$ and $P_{23}$ in eq.(\ref{phases})
 by applying the Sylvester
criterion, in order to yield positive eigenvalues in $P$.
Note that the Majorana phase, $\phi$, will not appear in
these expressions, and the only phase that will appear is the
Dirac phase, suppressed by the small value of $\theta_{13}$.
To be precise, if we substitute 
$\theta_{23}\simeq \pi/4$, $\theta_{12}\simeq \pi/6$
and we take into account that $\theta_{13}$ is small,
we obtain
\bea
P_{13}&\simeq&-|P_{13}|(1+ i~\sqrt{3}\sin\delta \sin \theta_{13}), 
\nonumber \\
P_{23}&\simeq&|P_{23}|(1+ 2i~\sqrt{3}\sin\delta \sin \theta_{13}).
\eea
Therefore,  in the limit where $|P_{12}|\ll|P_{13}|,|P_{23}|$,
all the elements in $P$ are expected to be real to a good
approximation, and accordingly the contribution
to the electric dipole moments from the slepton
parameters is expected to be very small. Note 
also that requiring positive eigenvalues for $P$ requires
that $P_{13}$ is negative and $P_{23}$ positive.

The high-energy parameters can be
easily reconstructed from the general formulas
in eq.(\ref{reconstruction}),  where the relevant elements 
in the matrix $Q$ read:
\bea
Q_{22}&\simeq& |P_{13}| \frac{|U_{23}|^2}{|U_{11}||U_{31}|}
+|P_{23}| \frac{|U_{13}|^2}{|U_{21}||U_{31}|},
\nonumber \\
Q_{33}&\simeq& |P_{13}| \frac{|U_{22}|^2}{|U_{11}||U_{31}|}
+|P_{23}| \frac{|U_{12}|^2}{|U_{21}||U_{31}|},
\nonumber \\
Q_{23}&\simeq& -|P_{13}| \frac{U^*_{22} U_{23}}{|U_{11}||U_{31}|}
-|P_{23}| \frac{U^*_{12}U_{13}}{|U_{21}||U_{31}|}.
\eea

To show the analytical results, we will limit ourselves to the  case
where $\theta_{23}\simeq \pi/4$, $\theta_{12}\simeq \pi/6$, 
and  $\theta_{13}\simeq 0$, as suggested by data.
Depending on the values of the
remaining non-vanishing parameters, we 
can distinguish two limits: $|P_{13}|\ll |P_{23}|$
and $|P_{23}|\ll |P_{13}|$. The structure 
of the $P$-matrix is different in each case, and reads:
\bea
P\simeq |P_{23}|
\pmatrix{
\lambda/\sqrt{6} & 0 &-\lambda \cr
0 & 1 & 1 \cr
-\lambda & 1 & 1},{~ \rm for~} \lambda= \frac{|P_{13}|}{|P_{23}|}\ll1,  \\
P\simeq |P_{13}|
\pmatrix{
1/\sqrt{6} & 0 & -1 \cr
0 &\lambda &\lambda \cr
-1 & \lambda & \sqrt{6}},{~ \rm for~}  \lambda= \frac{|P_{23}|}{|P_{13}|}\ll1. 
\eea

In this approximation there are no phases in $P$, 
although there could be phases in the neutrino mass matrix.
Let us analyze first the case where the neutrino mass matrix
is also real, and later on,  the general case allowing
complex parameters.}

\subsection{Real case}

The reconstruction of the high-energy parameters
in terms of the low energy parameters is 
straightforward using eq.(\ref{reconstruction-real}).
The reconstructed high energy parameters  in each limit read:
\begin{itemize}
\item{$|P_{13}|\ll |P_{23}|$
\bea
M_1&\simeq& 2\sqrt{\frac{2}{3}} \frac{|P_{13}| }{m_2} 
\langle H^0_u \rangle^2, \nonumber \\
M_2&\simeq& \frac{2 |P_{23}|}{m_3} \langle H^0_u \rangle^2,
\nonumber \\
\cos 2z &\simeq& 1.
\eea
so that the reconstructed Yukawa coupling is:
\bea
{\bf Y}_{\nu}\simeq \sqrt{|P_{23}|}
\pmatrix{\sqrt{\frac{|P_{13}|}{\sqrt{6}|P_{23}|}} &
\sqrt{\sqrt{\frac{3}{8}}\frac{|P_{13}|}{|P_{23}|}} &
-\sqrt{\sqrt{\frac{3}{8}}\frac{|P_{13}|}{|P_{23}|}} \cr
-\frac{|P_{13}|}{2|P_{23}|} & 1 & 1}.
\label{rec-Yukawa1}
\eea

It is interesting that in this limit
the lightest right-handed mass is essentially determined
by the rate for the process $\tau \rightarrow e \gamma$,
while the heaviest one, by the process 
$\tau \rightarrow \mu \gamma$. On the other
hand, the complex angle
in $R$ is such that in this case $\cos 2z$ 
is very close to one, independently
of the values of $|P_{13}|$ and $|P_{23}|$, as long
as $|P_{13}|\ll|P_{23}|$. 
}
\item{
$|P_{23}|\ll |P_{13}|$
\bea
M_1&\simeq& \frac{8 |P_{23}| }{3m_2+4m_3}\langle H^0_u \rangle^2,
\nonumber \\
M_2&\simeq& \frac{(3m_2+4m_3)|P_{13}|}{\sqrt{6}m_2 m_3}
\langle H^0_u \rangle^2,
\nonumber \\
\cos 2z &\simeq& \frac{3m_2-4m_3}{3m_2+4m_3},
\eea
for the case in which the light neutrinos have the same CP
parities, {\it i.e.} when $\phi=0$, and
\bea
M_1&\simeq& \frac{8 |P_{23}| }{-3m_2+4m_3}\langle H^0_u \rangle^2,
\nonumber \\
M_2&\simeq& \frac{(-3m_2+4m_3)|P_{13}|}{\sqrt{6}m_2 m_3}
\langle H^0_u \rangle^2,
\nonumber \\
\cos 2z &\simeq& \frac{3m_2+4m_3}{3m_2-4m_3},
\eea
when they have opposite CP parities, $\phi=\pi$.

The reconstructed Yukawa coupling is in this case:
\bea
{\bf Y}_{\nu}\simeq \sqrt{\sqrt{6}|P_{13}|}
\pmatrix{\sqrt{\frac{\sqrt{6}|P_{23}|}{|P_{13}|}}\frac{m_2}{3m_2+4m_3} &
\sqrt{\frac{|P_{23}|}{\sqrt{6}|P_{13}|}} &
\sqrt{\frac{|P_{23}|}{\sqrt{6}|P_{13}|}}\frac{-3m_2+4m_3}{3m_2+4m_3} 
\cr -\frac{1}{\sqrt{6}} & \frac{\sqrt{6}m_2}{3m_2+4m_3}
\frac{|P_{23}|}{|P_{13}|} & 1},
\label{rec-Yukawa2}
\eea
when $\phi=0$ and
\bea
{\bf Y}_{\nu}\simeq \sqrt{\sqrt{6}|P_{13}|}
\pmatrix{-\sqrt{\frac{\sqrt{6}|P_{23}|}{|P_{13}|}}\frac{m_2}{-3m_2+4m_3} &
\sqrt{\frac{|P_{23}|}{\sqrt{6}|P_{13}|}} &
\sqrt{\frac{|P_{23}|}{\sqrt{6}|P_{13}|}}\frac{3m_2+4m_3}{-3m_2+4m_3} 
\cr -\frac{i}{\sqrt{6}} & \frac{-i \sqrt{6}m_2}{-3m_2+4m_3}
\frac{|P_{23}|}{|P_{13}|} & i},
\label{rec-Yukawa3}
\eea
when $\phi=\pi$.

In the case with $|P_{23}|\ll |P_{13}|$, 
the behaviour is opposite to the previous one: 
the lightest right-handed mass is determined by 
$\tau \rightarrow \mu \gamma$, and the heaviest  by
$\tau \rightarrow e \gamma$. In this limit, $\cos 2z$
 is also  independent of  $|P_{13}|$ and $|P_{23}|$, and
takes a negative value, $\cos 2z\simeq -0.75$, for
$\phi=0$ and positive, $\cos 2z\simeq -1.33$, for $\phi=\pi$.
}
\end{itemize}

\begin{figure}[t!]
  \centerline{
    \scalebox{0.33}{\includegraphics{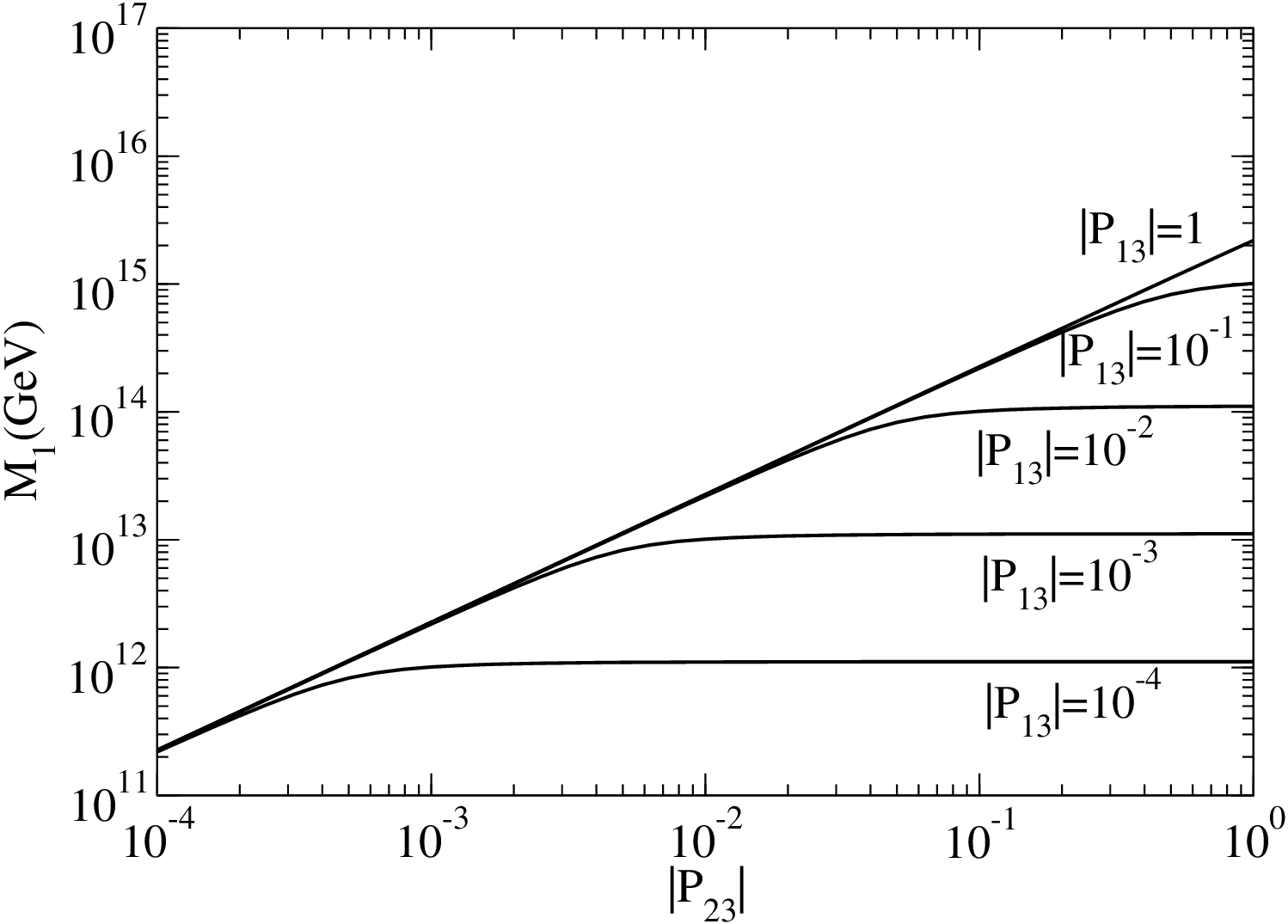}}
    ~
    \scalebox{0.33}{\includegraphics{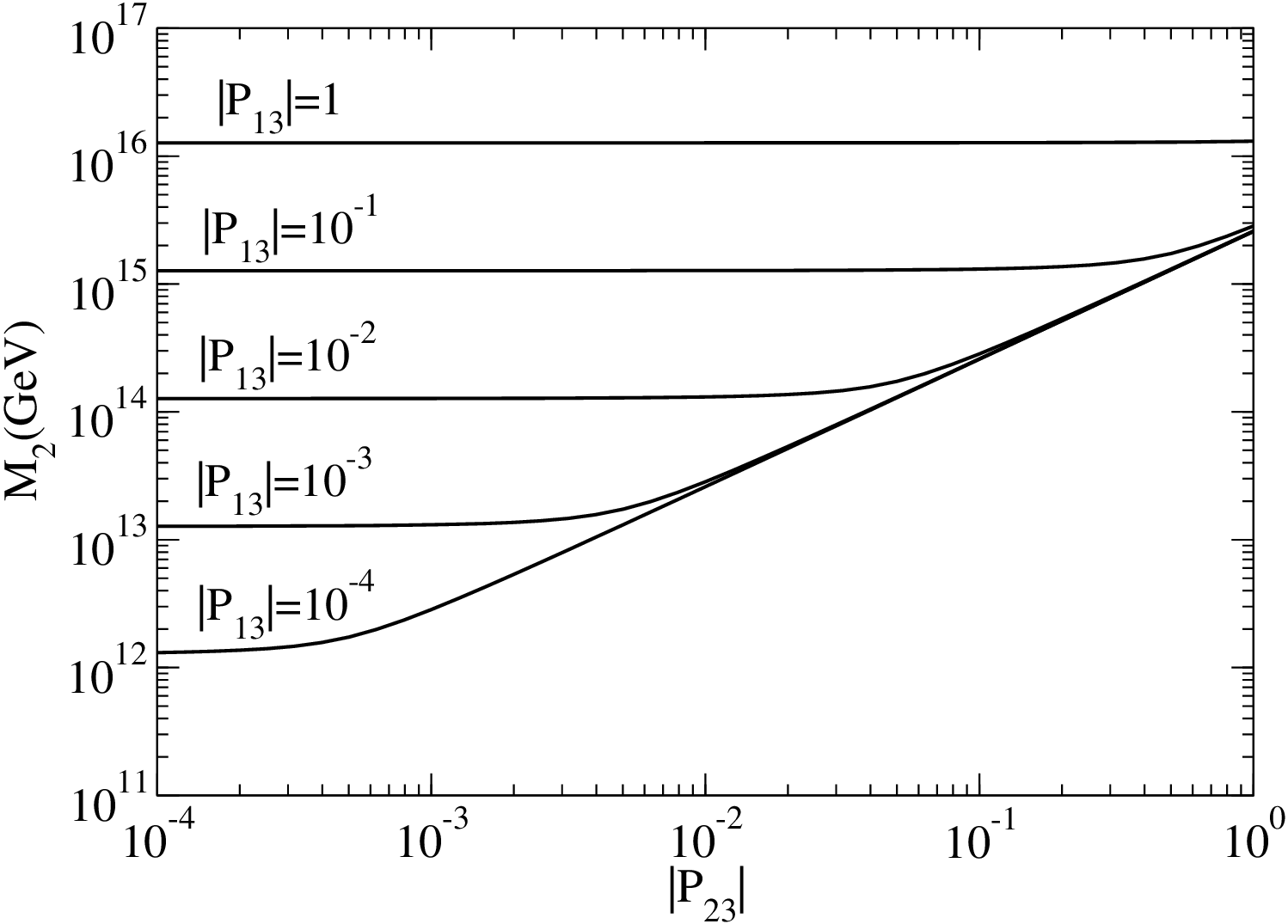}}
  } 
\vspace{.9cm}
 \centerline{
    \scalebox{0.33}{\includegraphics{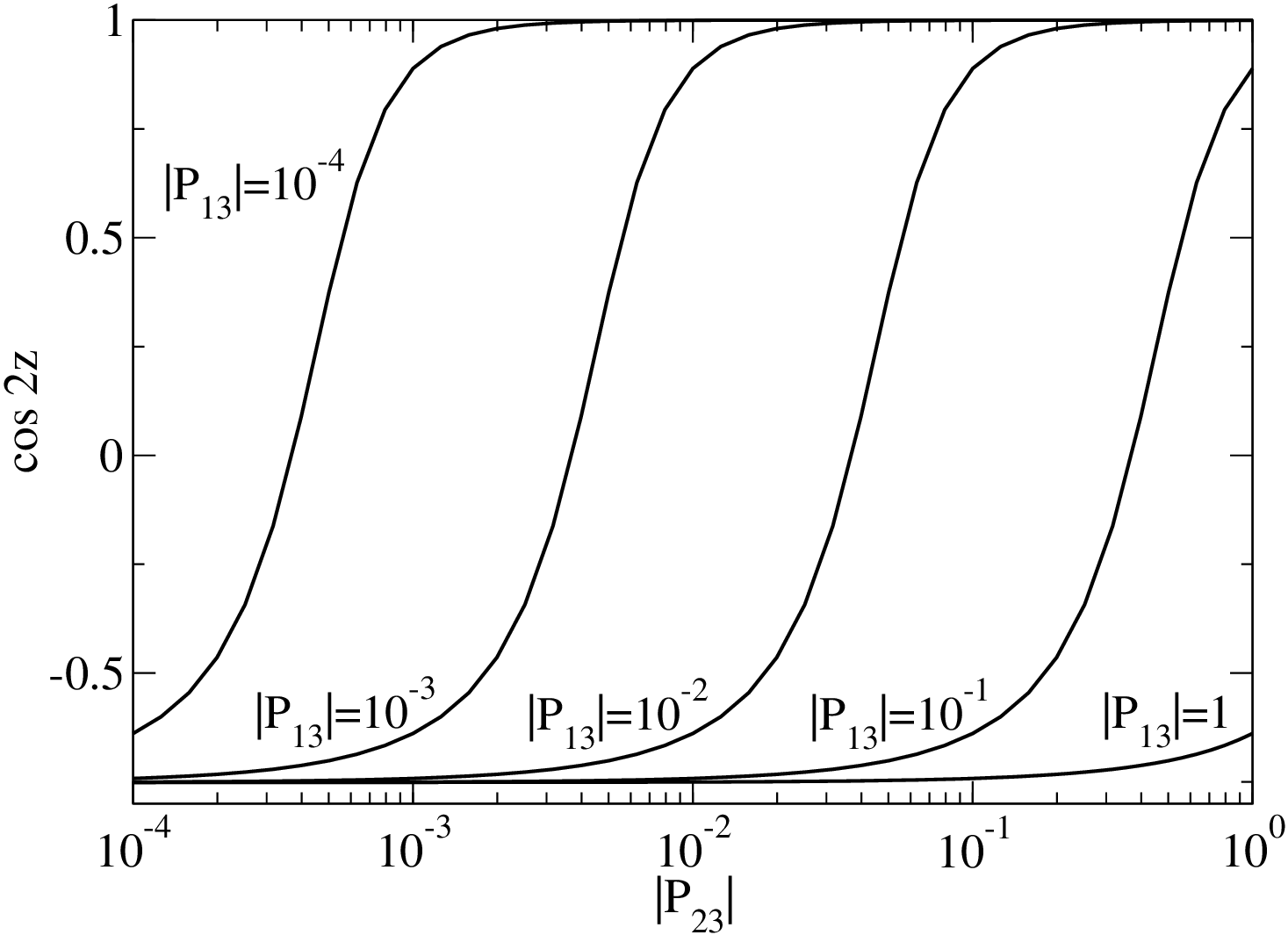}}
  }
  \caption{\small Reconstructed right-handed neutrino masses
and parameter $z$ in the matrix $R$, for different values
of $|P_{13}|$ and $|P_{23}|$, in the 
limit $|P_{12}|\ll  |P_{13}|, |P_{23}|$. Here, we have
taken $\theta_{23}\simeq \pi/4$, $\theta_{12}\simeq \pi/6$, 
$\theta_{13}\simeq 0$ and $\phi\simeq 0$ (so that all 
the parameters involved in the reconstruction procedure
are real to a good approximation).}
  \label{fig-real}
\end{figure}

The numerical results for the case with $\phi=0$
are shown in Fig.\ref{fig-real} 
for different values of $|P_{13}|$ and $|P_{23}|$, 
where the two  limits can be clearly distinguished.
(Recall that for slepton masses of $\sim 400$GeV and 
$\tan\beta=10$, $|P_{13,23}|<0.09$, however, we 
show the results for  $|P_{13,23}|<1$ to allow for
larger slepton masses or smaller values of $\tan\beta$.)
In these figures, we have used the central
values for the measured masses and mixing angles 
in \cite{Maltoni:2004ei}, namely $m_2=8.9\times10^{-3}$eV,
$m_3=4.7\times 10^{-2}$eV, $\sin^2\theta_{12}=0.30$
and $\sin^2\theta_{23}=0.50$. These parameters have to be run
from the electroweak scale to the decoupling scale
\cite{running} and this introduces corrections smaller
than a 60\% on the reconstructed parameters. In addition,
the experimental error on the low energy parameters
introduces an indeterminacy on the reconstructed parameters
smaller than a factor of two, that will be reduced 
in forthcoming experiments. It is apparent from
this analysis that the observation of rare 
decays would be an important step
towards reconstructing the complete Lagrangian
of the 2RHN model.

\subsection{Complex case, $\theta_{13}=0$}

For the complex case, we will show first the results 
for $\theta_{13}=0$, so that the high-energy 
parameters will depend only on the Majorana phase $\phi$ 
(recall that in this case $P$ is real with a good approximation). 
Later on, we will discuss the situation with non-vanishing  
$\theta_{13}$. On the other hand, 
for the atmospheric and solar angles, we maintain the
experimentally favoured values $\theta_{23}\simeq \pi/4$,
$\theta_{12}\simeq \pi/6$.
The analytical expressions for the high energy parameters are
given in this case by:
\begin{itemize}
\item{
$|P_{13}|\ll |P_{23}|$
\bea
M_1&\simeq& 2\sqrt{\frac{2}{3}} \frac{|P_{13}| }{m_2} \langle H^0_u \rangle^2,
\nonumber \\
M_2&\simeq& \frac{2 |P_{23}| }{m_3}\langle H^0_u \rangle^2,
\nonumber \\
\cos 2z &\simeq& 1.
\eea
Therefore, the Yukawa coupling is:
\bea
{\bf Y}_{\nu}\simeq \sqrt{|P_{23}|}
\pmatrix{\sqrt{\frac{|P_{13}|}{\sqrt{6}|P_{23}|}} e^{i \phi/2}&
\sqrt{\sqrt{\frac{3}{8}}\frac{|P_{13}|}{|P_{23}|}} e^{i \phi/2}&
-\sqrt{\sqrt{\frac{3}{8}}\frac{|P_{13}|}{|P_{23}|}}e^{i \phi/2} \cr
-\frac{|P_{13}|}{2|P_{23}|} & 1 & 1}.
\eea
Note that although in this limit the right-handed masses and $\cos 2z$
do not depend on the Majorana phase, the Yukawa coupling does. The 
dependence on the phase results from the phase in the leptonic
mixing matrix, $U$, that enters in the parametrization
of the Yukawa coupling, eq.(\ref{yukawa}).
Similarly to the real case, the lightest 
right-handed mass is essentially determined by
the rate for the process $\tau \rightarrow e \gamma$,
while the heaviest one, by the process 
$\tau \rightarrow \mu \gamma$. 
}
\item{$|P_{23}|\ll |P_{13}|$
\bea
M_1&\simeq& \frac{8 |P_{23}|\langle H^0_u \rangle^2}
{\sqrt{9m^2_2+16m^2_3+24 m_2 m_3 \cos\phi}} ,
\nonumber \\
M_2&\simeq& \frac{|P_{13}|\sqrt{9 m^2_2
+16 m^2_3+24 m_2 m_3 \cos \phi}}{\sqrt{6}m_2 m_3}\langle H^0_u \rangle^2,
\nonumber \\
\cos 2z &\simeq& \frac{3m_2e^{i\phi}-4m_3}{3m_2e^{i\phi}+4m_3},
\eea
so that the reconstructed Yukawa coupling is:
\bea
{\bf Y}_{\nu}\simeq \sqrt{\sqrt{6}\frac{\Delta}{\Delta^*}|P_{13}|} 
\pmatrix{\sqrt{\frac{\sqrt{6}|P_{23}|}{|P_{13}|}}\frac{m_2 e^{i\phi}}
{|\Delta|^2} &
\sqrt{\frac{|P_{23}|}{\sqrt{6}|P_{13}|}}\frac{\Delta^*}{\Delta} &
\sqrt{\frac{|P_{23}|}{\sqrt{6}|P_{13}|}}\frac{-3m_2 e^{i\phi} +4m_3}
{|\Delta|^2}
\cr -\frac{e^{i\phi/2}}{\sqrt{6}} & \frac{\sqrt{6}m_2e^{-i\phi/2}}
{\Delta^2}\frac{|P_{23}|}{|P_{13}|} & e^{i\phi/2}},
\eea
where $\Delta=\sqrt{3 m_2 e^{-i \phi}+4m_3}$.
Contrary to the previous limit, here the right-handed masses
and $R$ do depend on the Majorana phase. Concerning which
processes are relevant to determine which parameter,
the conclusion is analogous to the real case: 
the lightest right-handed mass is essentially determined
by the process
$\tau \rightarrow \mu \gamma$, and the heaviest by
$\tau \rightarrow e \gamma$.
}
\end{itemize}

\begin{figure}[t!]
  \centerline{
    \scalebox{0.33}{\includegraphics{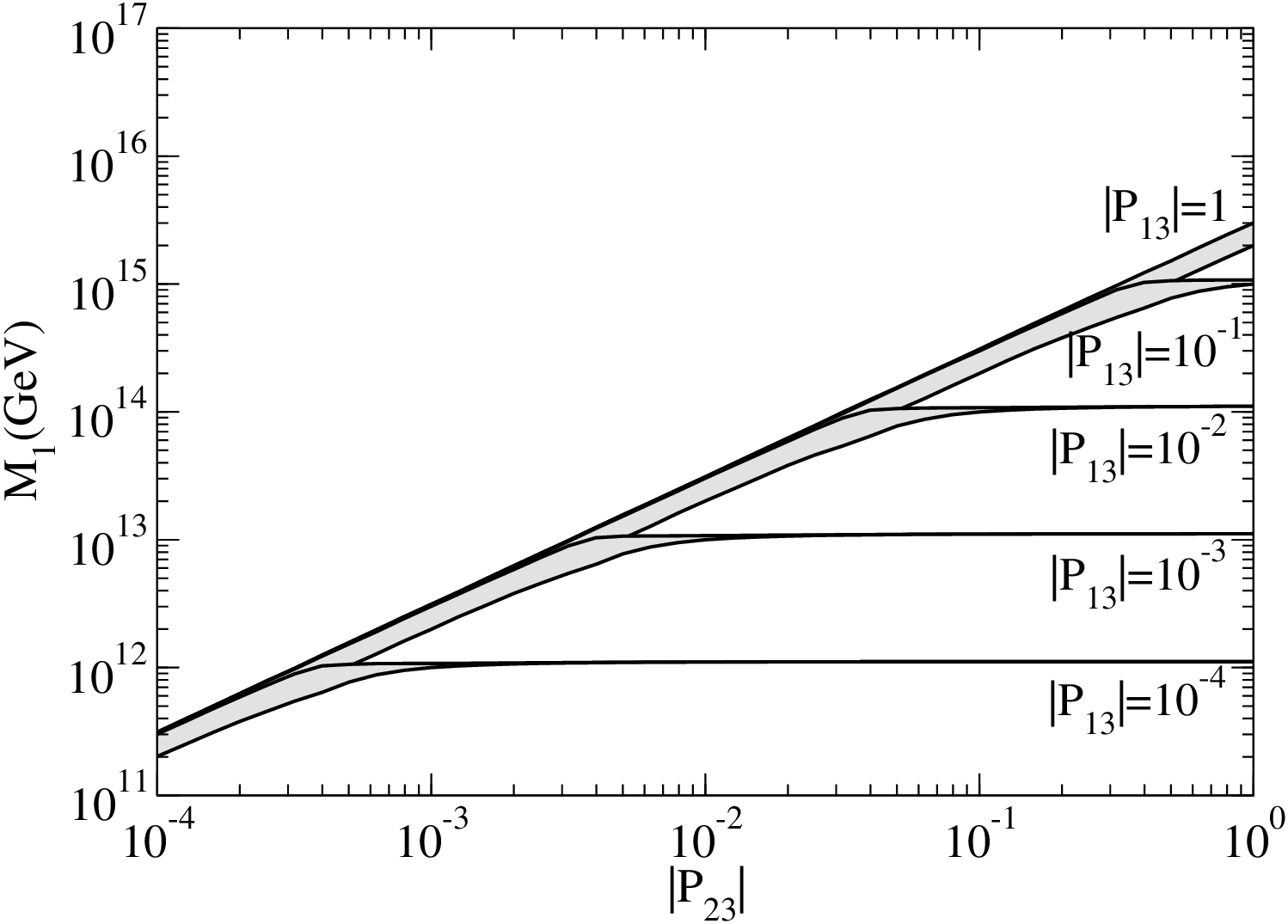}}
    ~
    \scalebox{0.33}{\includegraphics{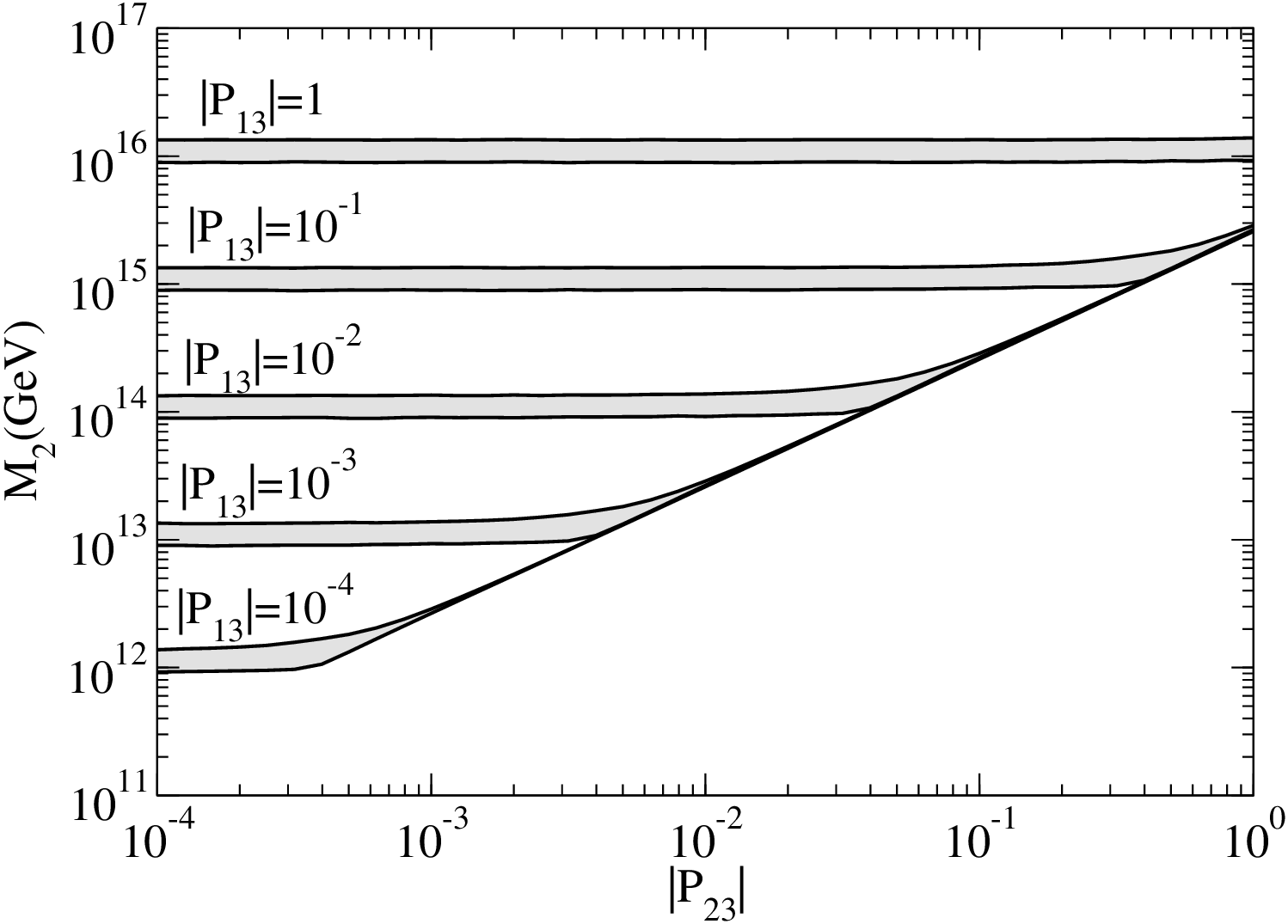}}
  } 
\vspace{.9cm}
 \centerline{
    \scalebox{0.33}{\includegraphics{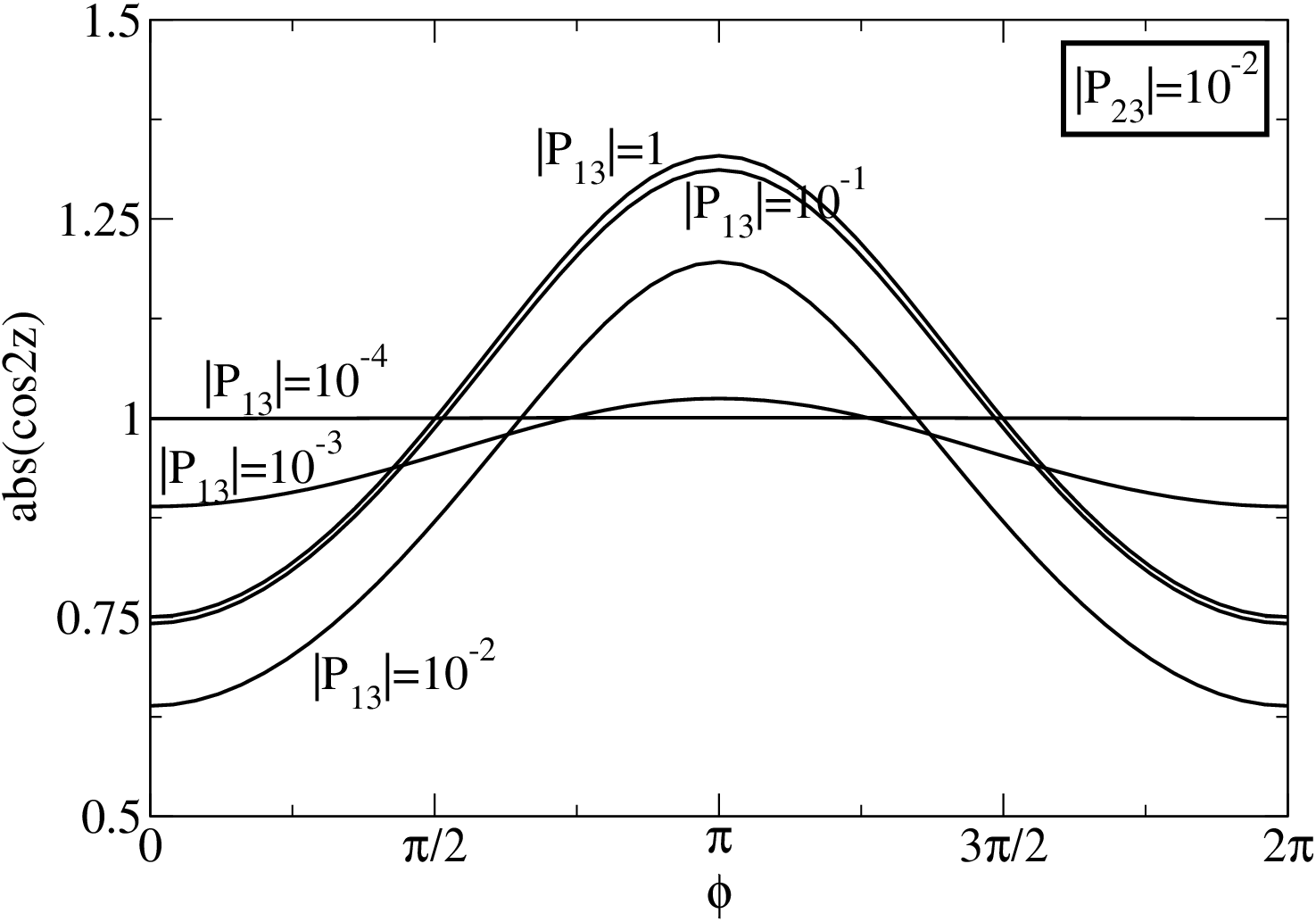}}
    ~
    \scalebox{0.33}{\includegraphics{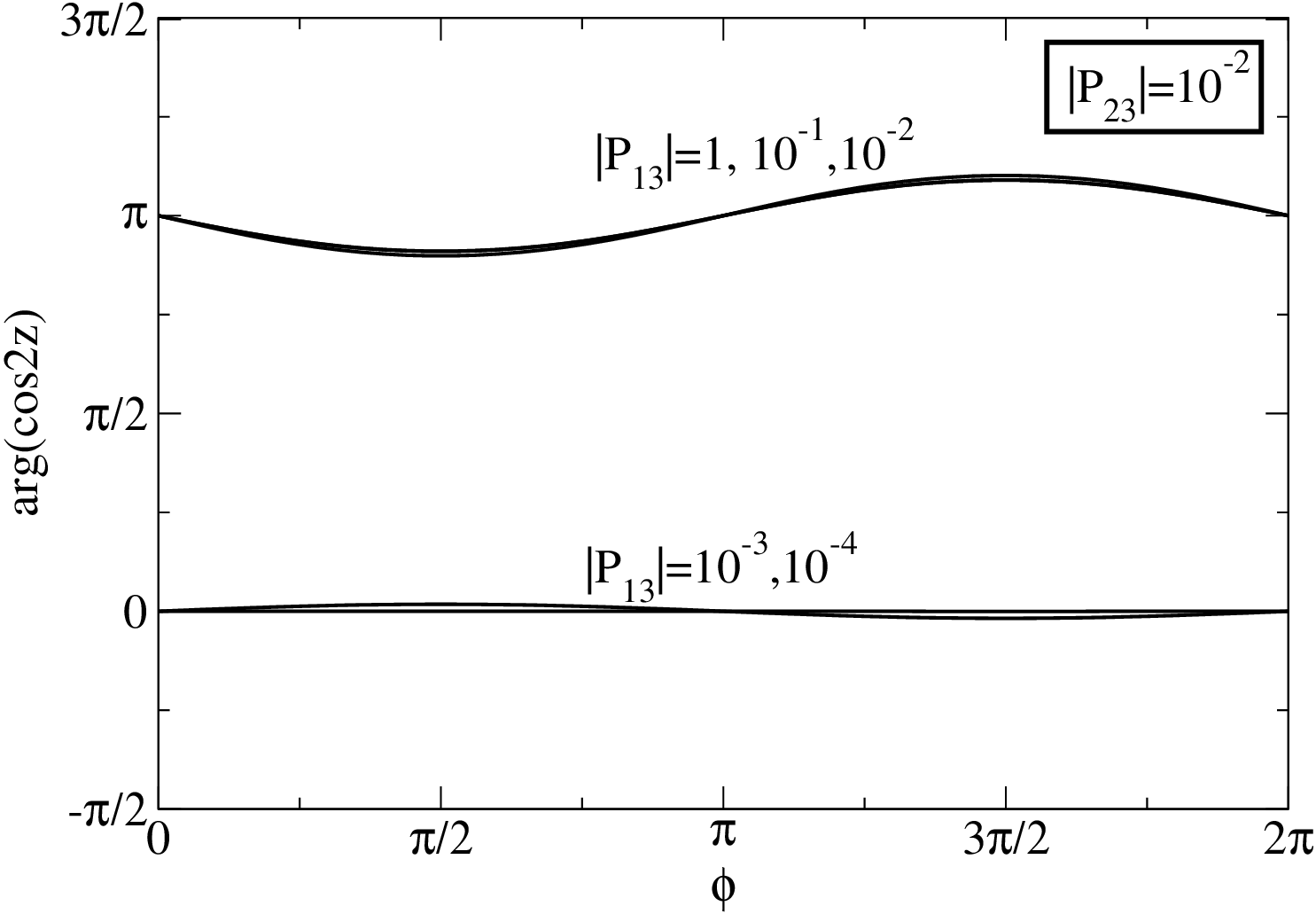}}
  }
  \caption{\small The same as Fig.\ref{fig-real}, but sampling
over different values for $\phi$. The shaded areas 
represent the regions at 2$\sigma$ from the main value. 
For $\cos 2z$ we plot both the absolute value and the
argument as a function of the Majorana phase $\phi$,
for fixed $|P_{23}|=10^{-2}$ and different values
of $|P_{13}|$.
}
  \label{fig-complex}
\end{figure}

The numerical results can be found in Fig.\ref{fig-complex}. We 
see that when $|P_{13}|\ll |P_{23}|$ the correlation
between the right-handed masses and $|P_{13}|$ or $|P_{23}|$
is very tight, allowing a very precise reconstruction of the
high-energy parameters. However, when  $|P_{23}|\ll |P_{13}|$ 
the reconstruction is more complicated, and the precise determination
of the high-energy parameters would require 
the measurement of the Majorana phase. In  Fig.\ref{fig-complex}
we have sampled $\phi$ between 0 and $2\pi$ and show the
regions at 2$\sigma$ from the main value. These regions
are fairly narrow, so even without knowing the 
Majorana phase, it could be possible
to reconstruct the right-handed masses from rare decays,
up to a factor of two.  

On the other hand, for the numerical value of $\cos 2z$
we show both the absolute value and the argument 
as a function of the Majorana phase, $\phi$,
for fixed $|P_{23}|=10^{-2}$ and for different values
of $|P_{13}|$. It can be checked that the prediction
for $\cos 2z$ depends on the ratio $|P_{13}|/|P_{23}|$,
so the results for other values of $|P_{23}|$ could
be read easily from this figure. We find that when 
$|P_{13}|\ll |P_{23}|$ both the absolute value and
the argument of $\cos2z$ are not very sensitive to
$\phi$. However, despite the matrix $R$ depends weakly on 
$\phi$, it is not possible to reconstruct the Yukawa coupling, 
due to the dependence of the Yukawa matrix with $\phi$
through the leptonic mixing matrix in eq.(\ref{yukawa}).
On the other hand, when  $|P_{23}|\ll |P_{13}|$,  the absolute
value of $\cos2z$ depends strongly on $\phi$. Therefore,
in both cases the reconstruction
of the Yukawa coupling would require the determination of the
Majorana phase, although it would not be necessary for a rough
reconstruction of the right-handed masses.

\subsection{Complex case, $\theta_{13}\neq0$}

\begin{figure}[t!]
  \centerline{
    \scalebox{0.33}{\includegraphics{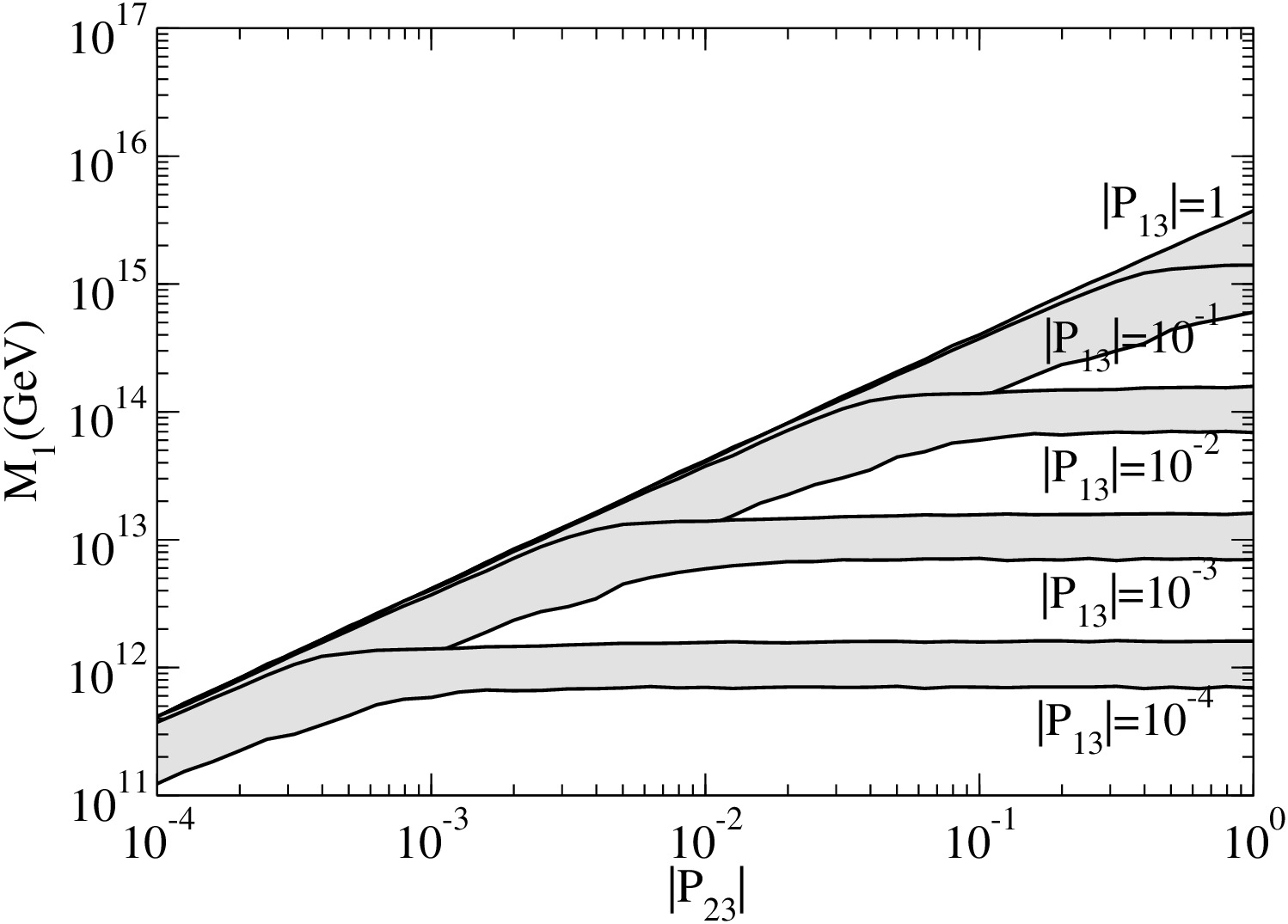}}
    ~
    \scalebox{0.33}{\includegraphics{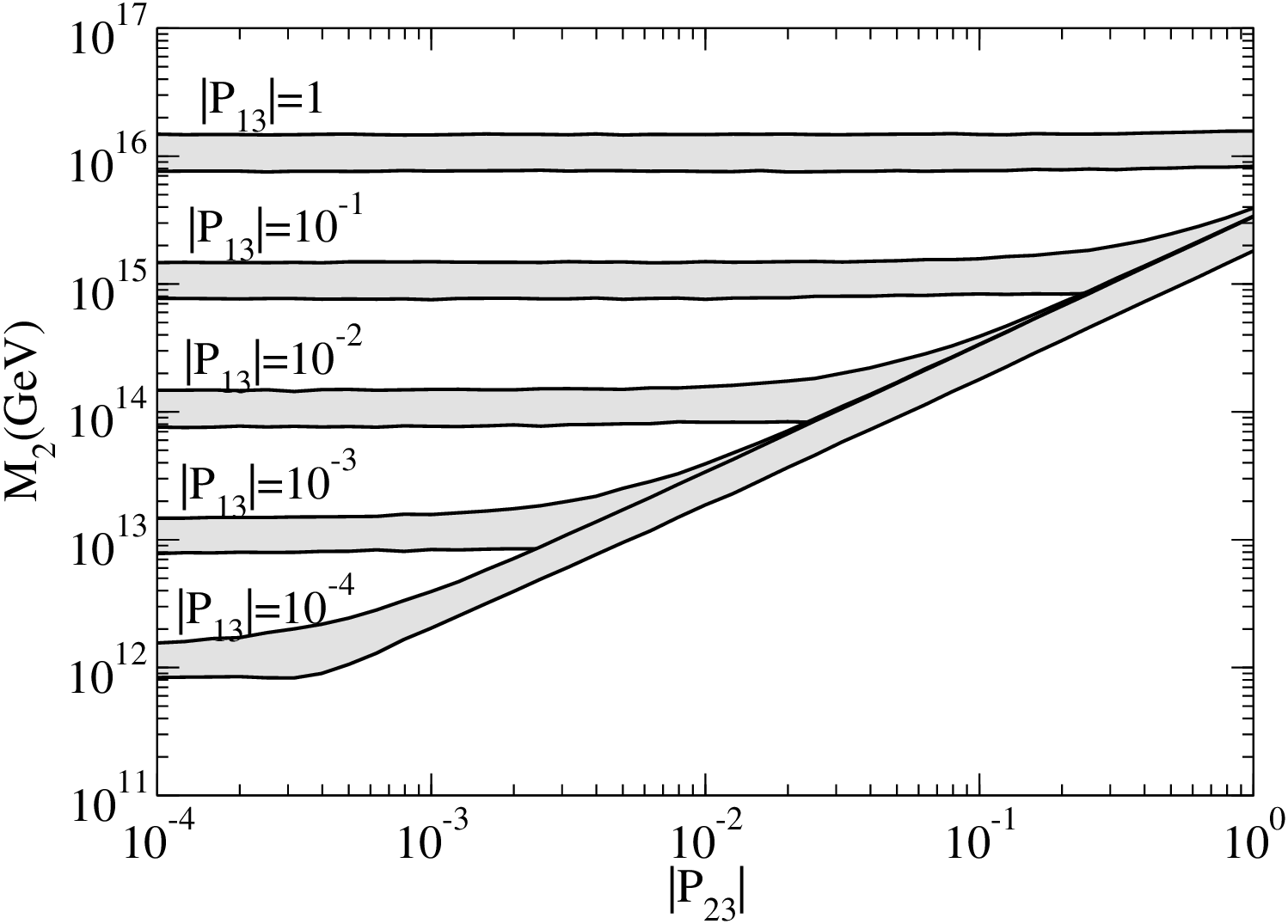}}
  } 
\vspace{.9cm}
 \centerline{
    \scalebox{0.33}{\includegraphics{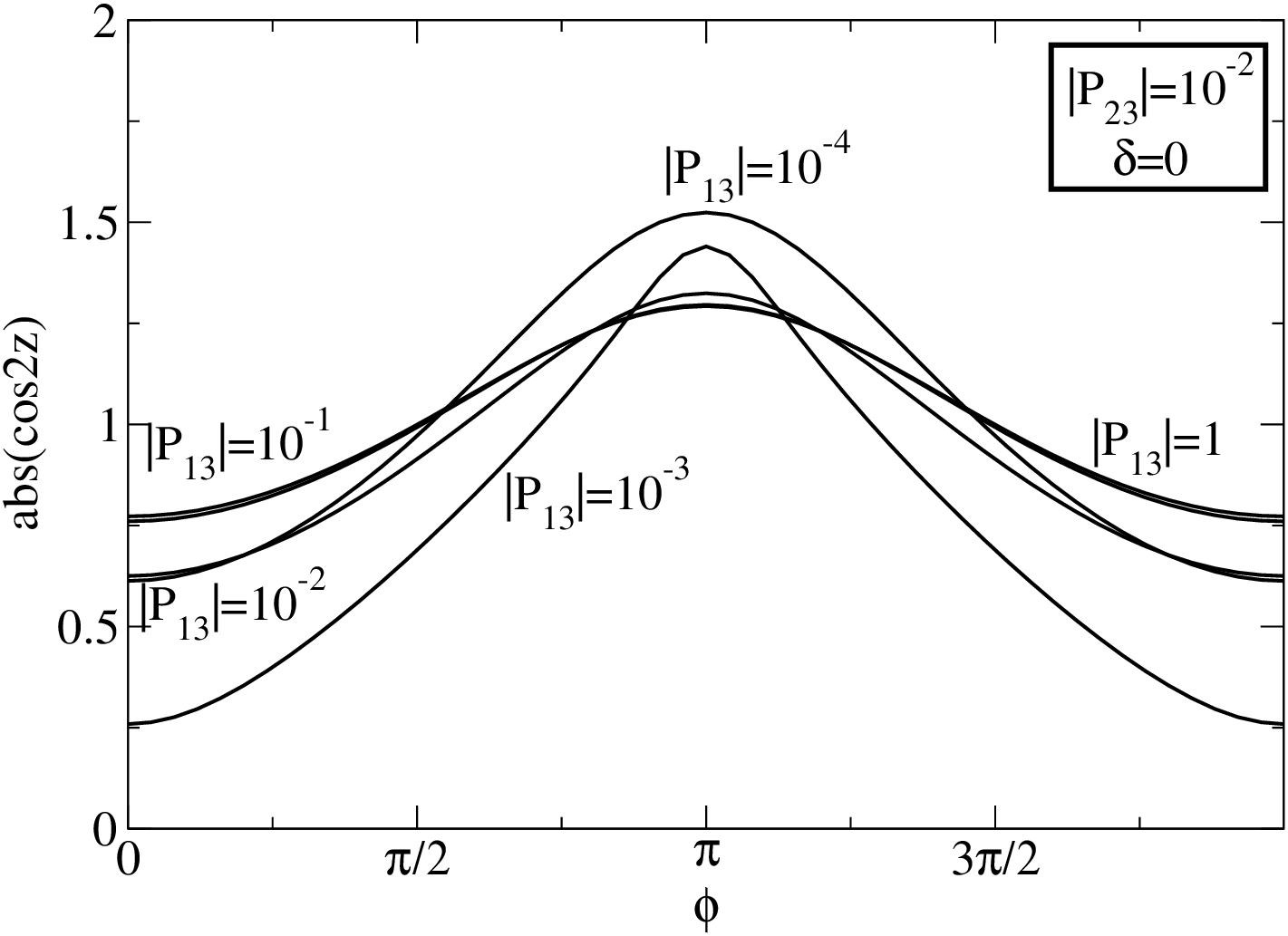}}
    ~
    \scalebox{0.33}{\includegraphics{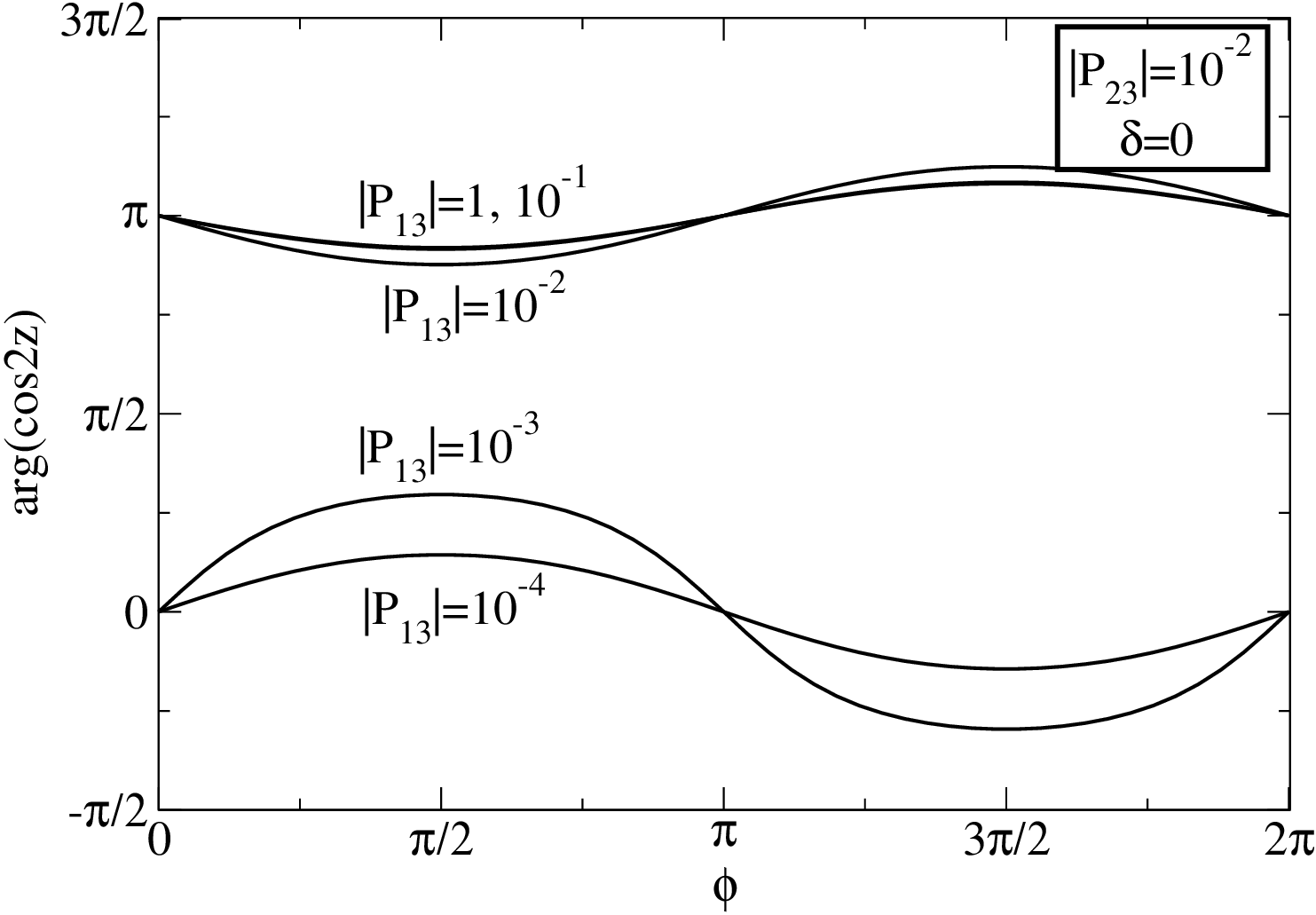}}
  }
\vspace{.9cm}
 \centerline{
    \scalebox{0.33}{\includegraphics{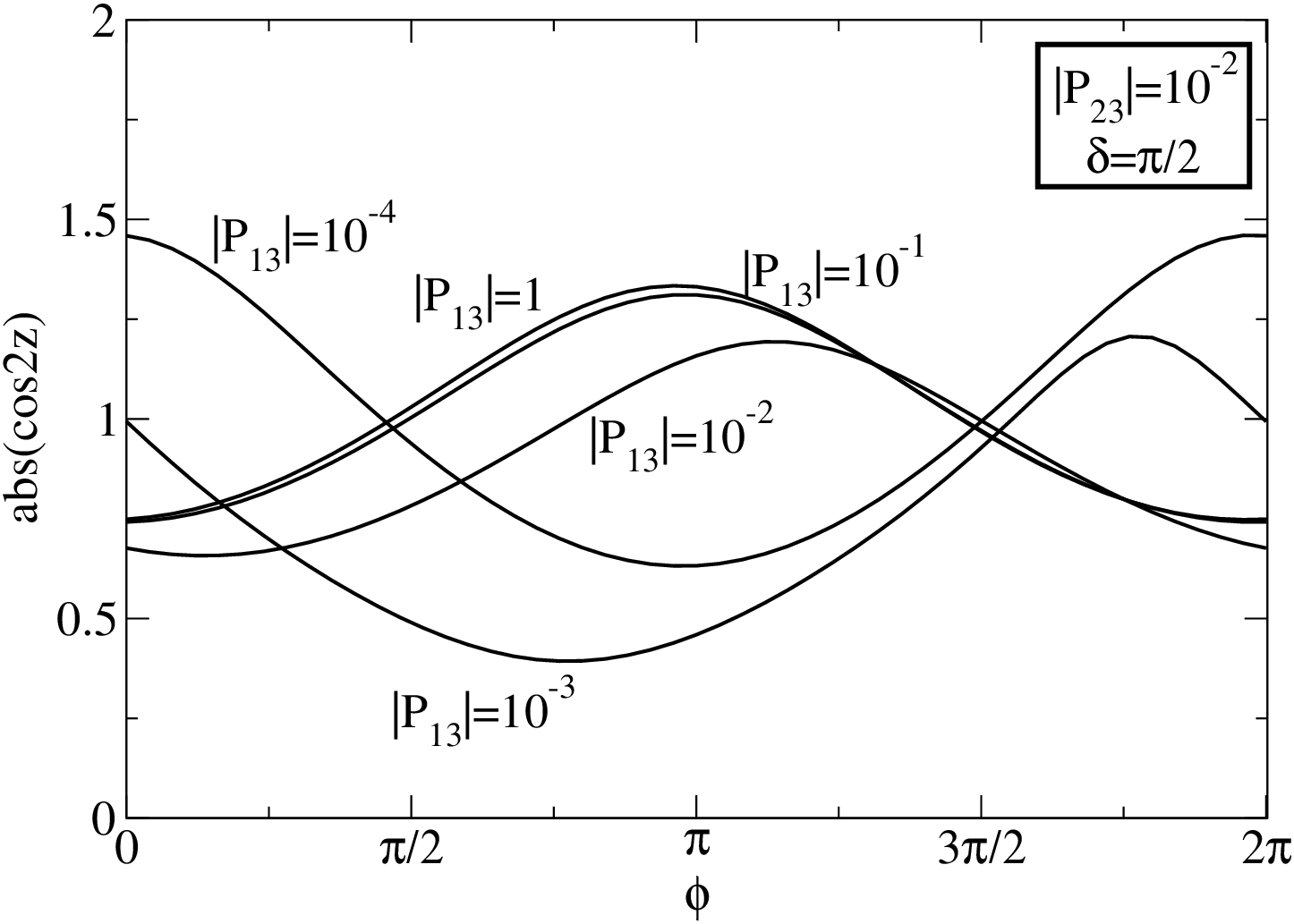}}
    ~
    \scalebox{0.33}{\includegraphics{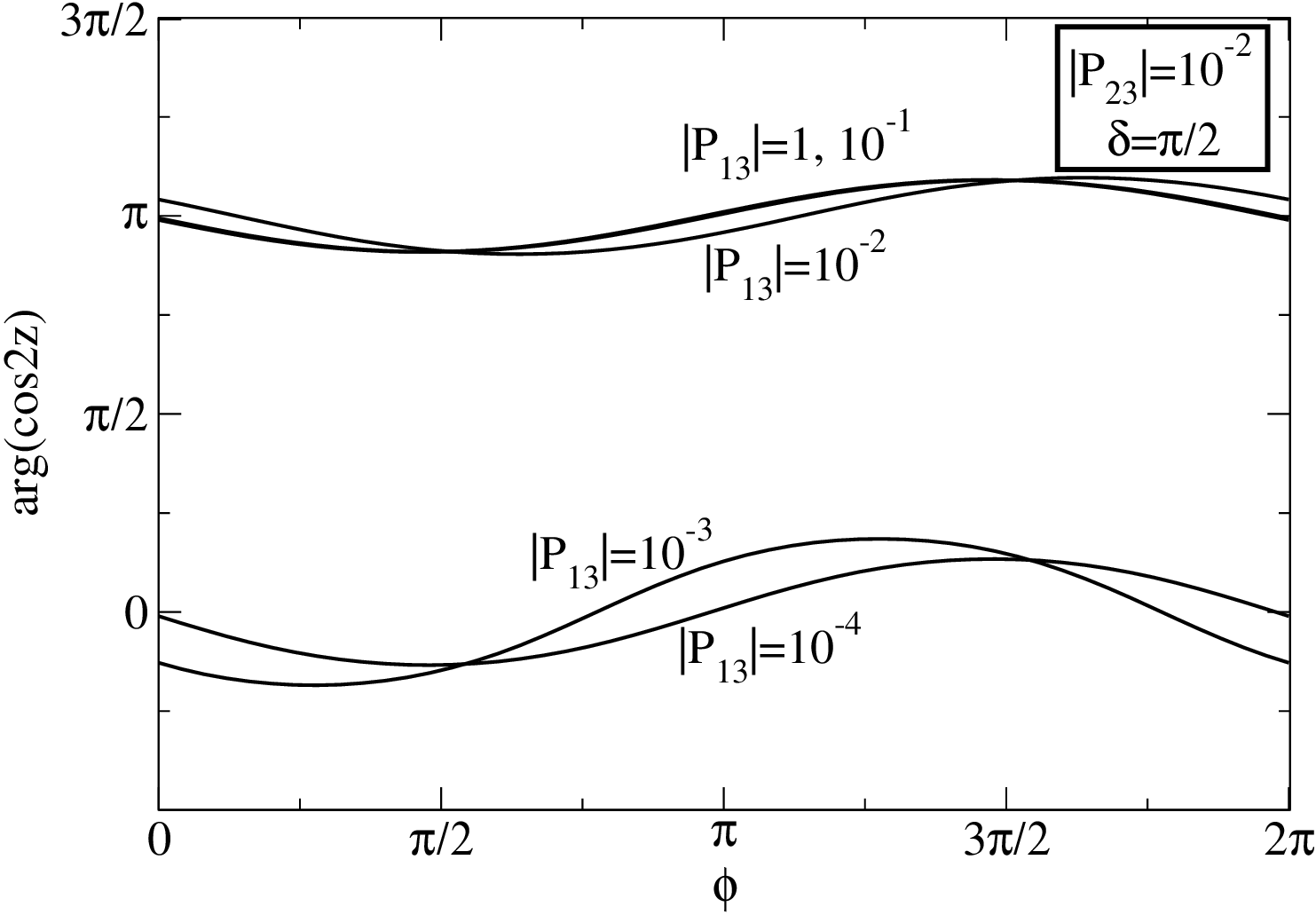}}
  }
  \caption{\small The same as Fig.\ref{fig-real}, but sampling
over different values for $\phi$ and $\delta$ for $\theta_{13}=0.1$.
The shaded areas represent the regions at 2$\sigma$ from the main value. 
For $\cos 2z$ we plot both the absolute value and the
argument as a function of the Majorana phase $\phi$,
for fixed $|P_{23}|=10^{-2}$ and different values
of $|P_{13}|$, and for two values of the phase $\delta$: 0 and $\pi/2$.}
  \label{fig-complex-13}
\end{figure}

Finally we show the results including the effects
of $\theta_{13}$ and the phase $\delta$. Although it is possible
to derive analytic expressions for the high-energy parameters
in the different limits, the expressions are very complicated
and difficult to analyze. Therefore, in this subsection we will
limit ourselves to show the numerical results, that
can be found in Fig.\ref{fig-complex-13}.

We find that in the whole parameter space, the measurement
of $\theta_{13}$, $\delta$ and $\phi$ would be desirable for
a precise determination of the right-handed masses. However,
even without knowing these parameters, the determination
of the rates for the rare lepton decays would allow the
reconstruction of the right-handed masses up to a factor
of three.

On the other hand, when $\theta_{13}$ is large, 
the reconstruction of the parameter $z$
necessarily requires the measurement of all the low-energy 
parameters. The situation is particularly critical when
$|P_{13}|\ll |P_{23}|$, since  $\cos2z$ is very sensitive
to the Dirac phase (when  $\theta_{13}$ is large). For
instance, when $\theta_{13}=0.1$, the absolute value
of $\cos2z$ can vary between 0.6 and 1.5 when $\delta$
and $\phi$ vary between 0 and $2\pi$. 
When $|P_{23}|\ll |P_{13}|$, the 
dispersion produced by the angle $\theta_{13}$ is 
smaller, between 0.8 and 1.3. Concerning the argument
of $\cos2z$ the range of values is smaller, although still
important,  particularly
in the limit $|P_{13}|\ll |P_{23}|$. 
Therefore, when  $\theta_{13}$ is large, 
it seems unavoidable the precise measurement of
$\theta_{13}$, $\delta$ and $\phi$ to determine
$\cos 2z$ and to reconstruct the Yukawa coupling.

\section{The 2RHN model as a limit of the 3RHN model}

Although the 2RHN model can explain all the neutrino
experiments, the 3RHN model is without any doubt much more
appealing from the theoretical point of view.
Nevertheless, there are some situations where the 
3RHN model can be well approximated
by a 2RHN model, both from the point of view of neutrino
masses, as from the point of view of radiative corrections,
parametrized by the matrix $P$. In these situations,
the procedure presented in this paper to reconstruct the
high-energy parameters would apply.

Let us discuss first the case of the neutrino mass matrix,
that in the basis where the right-handed neutrino mass
matrix is diagonal reads:
\begin{equation}
{\cal M}_{ij}=\frac{y_{1i}y_{1j}}{M_1}+\frac{y_{2i}y_{2j}}{M_2}
+\frac{y_{3i}y_{3j}}{M_3},
\end{equation}
where $y_{ij}=({\bf Y}_{\nu})_{ij}$.
Two right-handed neutrinos dominate the see-saw when
\bea
\frac{y_{1i}y_{1j}}{M_1}&\ll&\frac{y_{2i}y_{2j}}{M_2},\frac{y_{3i}y_{3j}}{M_3}
~~~~ {\rm or} \nonumber \\
\frac{y_{2i}y_{2j}}{M_2}&\ll&\frac{y_{1i}y_{1j}}{M_1},\frac{y_{3i}y_{3j}}{M_3}
~~~~  {\rm or}\nonumber  \\
\frac{y_{3i}y_{3j}}{M_3}&\ll&\frac{y_{1i}y_{1j}}{M_1},\frac{y_{2i}y_{2j}}{M_2}
~~~~  {\rm for~all~}i,j=1,2,3.
\eea
The most interesting cases are the first and the third. The first
one corresponds to the case in which the Yukawa couplings for
the first generation of right handed neutrinos are tiny,
$y_{1i}\ll y_{2i}, y_{3i}$, for $i=1,2,3$. If this
is the case, the radiative corrections are also dominated
by the same two right-handed neutrinos, the two heaviest ones. Therefore,
in this case the 3RHN model can be well approximated by
a 2RHN model, both from the point of view of neutrino masses
as of radiative corrections. Since the two relevant right-handed
neutrinos are the two heaviest ones, the corresponding
Yukawa couplings could be large, and the radiative corrections 
could be sizable.\footnote{In this scenario there would be no
relation whatsoever between leptogenesis and low energy observables:
leptogenesis would depend on the couplings of the lightest
right-handed neutrino, whereas neutrino and slepton parameters
would be determined by the couplings of the two heaviest right-handed
neutrinos.}

The third case corresponds to the situation where the mass of
the heaviest right-handed neutrino is much larger than the mass
of the other two,  $M_3\gg   M_1,M_2$  \cite{related}.
However, in general the heaviest right-handed neutrino
will produce sizable contributions to the radiative corrections. 
If this is the case, the 3RHN model could be reduced
to a 2RHN model only from the point of view of neutrino masses,
but not from the point of view of the radiative corrections.
Nevertheless, there are some circumstances in which the heaviest
right-handed neutrino indeed does not contribute to the radiative
corrections and does not leave any imprint in $P$, so that 
this matrix is only determined by the Yukawa couplings
of the two lightest generations of right-handed neutrinos.
If this occurs, the 3RHN model would also be well approximated
by a 2RHN model from the point of view of the radiative
corrections.  This situation arises for example
when the mass of the heaviest right-handed neutrino
is very close to the Planck mass,
although this possibility seems a bit contrived.

A more plausible situation arises in models with 
gauge mediated supersymmetry breaking \cite{Alvarez-Gaume:1981wy}.
So far, we have implicitly assumed that the boundary conditions for
the soft breaking terms are set at the Planck scale. 
However, if the mass of the messenger particles involved in the 
supersymmetry breaking mechanism 
is smaller than $M_3$ but larger than $M_2$, then the heaviest
right-handed neutrino would decouple at an energy larger than
the energy at which supersymmetry breaking is communicated
to the observable sector. Consequently, it would not
participate in the radiative corrections of the parameters
of the Lagrangian. If this is the case, only the two lightest right-handed
neutrinos would contribute to the radiative corrections
and to the neutrino mass generation, and therefore the 3RHN model
could be well approximated by a 2RHN model.

The experimental signature of
this scenario would be a light gravitino, although probably not 
ultra-light, since the  mass of the messenger has to be 
larger than the mass of the next-to-lightest right-handed neutrino,
that is expected to be rather large. The gravitino mass
in these scenarios can be estimated as  \cite{Giudice:1998bp}
\begin{equation}
m_{3/2}=\frac{F}{\sqrt{3} M_P}\sim  
\frac{\pi}{\sqrt{3}\alpha}\frac{M_{\rm mes}}{M_P}\;\widetilde m,
\end{equation}
where we have assumed vanishing cosmological constant
and $M_P=(8\pi G_N)^{-1/2}=2.4\times10^{18}$GeV is
the reduced Planck mass. 
In this formula, $F$ measures supersymmetry breaking in the messenger
sector, $M_{\rm mes}$ is the mass of the messenger particles
and  $\widetilde m \sim (\alpha/\pi) F/M_{\rm mes}$ is the
typical soft mass scale (recall that in gauge mediated
supersymmetry breaking scenarios gaugino masses are generated at
one loop and scalar masses at two loops). 
If only the heaviest right-handed neutrino decouples from
the radiative corrections, it has to happen that 
$M_2<M_{\rm mes}<M_3$. So, if the gravitino mass is measured,
a lower bound on $M_3$ would follow:
\bea
M_3\gsim\frac{ \sqrt{3}\alpha}{\pi} \frac{m_{3/2}}{\widetilde m} M_P.
\eea
Alternatively, if $M_2$
can be reconstructed from low energy data, a lower bound on
the gravitino mass would follow. Note that the precise
determination of $M_2$ requires the knowledge of the
cut-off scale, that is not known a priori. However
one could compute $M_2$ assuming that the cut-off is set
at the Planck scale instead of the actual messenger scale. 
Then, the value obtained for $M_2$ would be smaller
(see the Appendix for details) and therefore the
bound on the gravitino mass obtained in this way would
hold, although it would be more conservative than the 
actual bound.

Besides, when supersymmetry breaking is mediated
to the observable sector through gauge interactions 
by particles with a mass smaller than the Planck mass,
the soft terms are almost proportional to the identity at
the cut-off scale. Thus, any flavour changing effect or
mass splitting between sleptons or sneutrinos observed at low
energies would be entirely due to radiative corrections. 
On the other hand, CP violating effects 
could originate in other sectors, such as the neutralino
or the chargino sector, although there could be a contribution
from $P$ that might be disentangled.

Finally, if in addition the gravitino is the lightest
supersymmetric particle, some interesting cosmological
consequences would follow. For instance, the strong bounds
on the reheating temperature, $T_R\lsim 10^6$GeV for
$m_{3/2}\sim {\cal O}(1 {\rm TeV})$  \cite{Kawasaki:2004qu}, would be relaxed
so that it could be as large as $10^{11}$GeV
\cite{Ellis:1984er}. This has crucial
consequences for the leptogenesis mechanism, since the 
mass of the lightest right-handed could be compatible
with the constraints on the reheating temperature from
preventing gravitino overproduction 
\cite{Davidson:2002qv,Buchmuller:2004nz}.

\section{Conclusions}

The two right-handed neutrino model can explain all 
the neutrino oscillation experiments, but depends
on less parameters than the conventional three 
right-handed neutrino model. Therefore, the 
serious problem of the lack of predictivity 
of the conventional see-saw model is softened.
In this paper we have exploited this observation
to argue that the high-energy parameters of the 
two right-handed neutrino model could be 
reconstructed using just low energy experiments,
provided supersymmetry is discovered and 
some hypotheses are made about the structure
of the soft terms at the cut-off scale. 

To this end, we have proposed an alternative parametrization
of the two right-handed neutrino model just in terms
of low energy observables, namely the neutrino mass matrix
and the off-diagonal elements of the matrix  
${\bf Y}^{\dagger}_{\nu}{\bf Y}_{\nu}$, that is responsible
for the radiative corrections of the slepton parameters,
and that in particular induces rare leptonic decays.
We have discussed the present
information available on these parameters and the prospects
to improve our knowledge of them in the next few years. 
Except for the case of the Majorana phase, we find 
the prospects very encouraging.

We have presented an exact procedure that allows to reconstruct
the high-energy superpotential of the two right-handed neutrino
model in terms of the low energy parameters. We have applied
this procedure to a particular scenario with
$BR(\mu\rightarrow e \gamma)\ll 
BR(\tau\rightarrow (\mu,e) \gamma)$, and we have found
that in the case where all the parameters are real and
$\theta_{13}$ is small, 
the detection of the rare decays $\tau\rightarrow \mu \gamma$
and $\tau\rightarrow e \gamma$ would be a very important
step towards reconstructing the high-energy parameters.
We have also analyzed the impact of the phases in the 
leptonic mixing matrix on this conclusion, since 
they could be the worst determined among all the 
low energy parameters.
We have found that when the angle $\theta_{13}$ is large 
and the phases in the leptonic mixing matrix do not
vanish, the connection is more diffuse, although it could
still be possible to determine the right-handed masses
up to a factor of three. The reconstruction
of the neutrino Yukawa coupling is more complicated,
since in general would require the measurement of 
$\theta_{13}$, $\delta$ and $\phi$. 

Finally, we have argued that this procedure does not
apply only to the strict two right-handed neutrino model.
There are limits of the three right-handed neutrino model
that resemble a two right-handed
neutrino model to a good approximation, both from
the point of view of neutrino masses, as from the point
of view of radiative corrections. In these limits, the procedure
proposed in this paper applies.

\section*{Acknowledgments}
I would like to thank Alberto Casas, and especially
Sacha Davidson for very interesting discussions. I would
also like to thank the CERN Theory Division for hospitality 
during the last stages of this work.

\appendix{\section{Appendix}}

Throughout the paper we have used as parameters to reconstruct
the high-energy theory $P={\bf Y}^{\dagger}_{\nu}{\bf Y}_{\nu}$ 
and ${\cal M}= {\bf Y}_{\nu}^T {\rm diag}(M^{-1}_1,M^{-1}_2) 
{\bf Y}_{\nu}\langle H_u^0\rangle^2$. However, to disentangle
$P$ it is necessary to know not only the cut-off scale but also
the decoupling scales, that are not known a priori. Nevertheless,
the procedure can be applied recursively in order
to reconstruct the high-energy parameters. 

In this Appendix it will be shown that this 
recursive procedure could be avoided using as low energy
parameters of the 2RHN model
\bea
{\cal M}&=&{\bf Y}^{T}_{\nu}{\rm diag}(M^{-1}_1,M^{-1}_2)
{\bf Y}_{\nu} \nonumber \langle H^0_u \rangle^2, \\
P&=&{\bf Y}^{\dagger}_{\nu}{\rm diag}(\log\frac{\Lambda}{M_1},\log\frac{\Lambda}{M_2})
{\bf Y}_{\nu},
\label{appendix1}
\eea
where the second parameter is the combination that appears
in the leading-log approximation to the solution of
the renormalization group equations, and $\Lambda$ is the cut-off scale.

Defining the new parameters 
\bea
\widetilde {\bf Y}_{\nu}&=&{\rm diag}(\sqrt{\log\frac{\Lambda}{M_1}},
\sqrt{\log\frac{\Lambda}{M_2}}){\bf Y}_{\nu}, \nonumber \\
\widetilde M_1 &=& M_1 \log\frac{\Lambda}{M_1}, \nonumber \\
\widetilde M_2 &=& M_2 \log\frac{\Lambda}{M_1}
\label{appendix2}
\eea
and substituting in eq.(\ref{appendix1}), one finds that 
\bea
{\cal M}&=& \widetilde{\bf Y}^{T}_{\nu}
{\rm diag}(\widetilde M^{-1}_1,\widetilde M^{-1}_2)
\widetilde {\bf Y}_{\nu} \nonumber \langle H^0_u \rangle^2, 
\nonumber \\
P&=&\widetilde {\bf Y}^{\dagger}_{\nu}\widetilde {\bf Y}_{\nu}.
\eea
Therefore, we can apply the general formulas 
eq.(\ref{reconstruction}) to solve for the 
parameters $\widetilde{\bf Y}_{\nu}$, $\widetilde M_1$
and $\widetilde M_2$. Finally, inverting eq.(\ref{appendix2})
we can reconstruct the actual high-energy parameters
${\bf Y}_{\nu}$, $ M_1$ and $ M_2$. Note that there are
always two solutions for $M_1$ and $M_2$, one larger than
$\Lambda/e$ and the other smaller. 

\begin{figure}[t!]
 \centerline{
    \scalebox{0.33}{\includegraphics{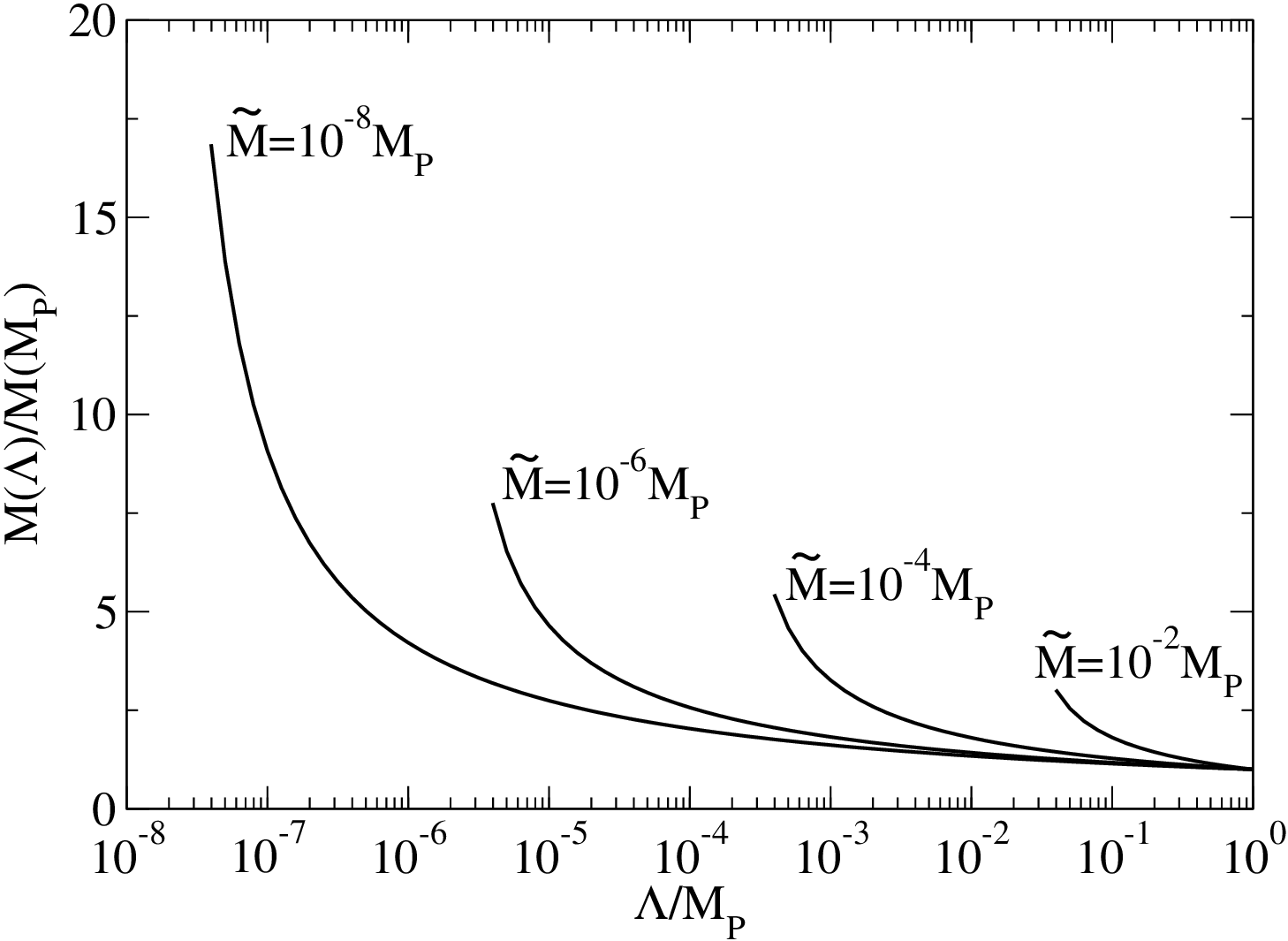}}
  }
  \caption{\small Relation between the right-handed neutrino
mass reconstructed using the actual cut-off scale
$\Lambda$, and reconstructed using the Planck mass as 
cut-off, for different values of $\widetilde M$
(see details in the Appendix).}
  \label{fig-appendix}
\end{figure}

From eq.(\ref{appendix2}) it is possible to estimate
the impact of our ignorance of the cut-off scale on the
reconstructed values of the right-handed masses. From the
reconstruction procedure it is always possible to compute
$\widetilde M_1$ and $\widetilde M_2$, and from eq.(\ref{appendix2}),
the actual masses $M_1$ and $M_2$, that would
depend on the cut-off scale. If we set the cut-off
scale at the Planck mass, we would obtain $M_1(M_P)$ and $M_2(M_P)$.
However, it could happen that the actual cut-off
scale is smaller than the Planck mass, for
instance in models with gauge mediated supersymmetry breaking;
in that case we would obtain $M_1(\Lambda)$ and $M_2(\Lambda)$.
In Fig.\ref{fig-appendix} we compare the values 
of the masses computed assuming that the cut-off 
is set at the Planck mass or at a different
cut-off, for fixed  $\widetilde M_1$ or  $\widetilde M_2$.
We find that taking as the cut-off the Planck mass instead
of an intermediate scale underestimates the  
value of the reconstructed right-handed masses. 
Nevertheless, the error made is usually smaller than one order
of magnitude.

\end{document}